\documentclass[sn-mathphys-num]{sn-jnl}

\usepackage{graphicx}%
\usepackage{multirow}%
\usepackage{amsmath,amssymb,amsfonts}%
\usepackage{amsthm}%
\usepackage{mathrsfs}%
\usepackage[title]{appendix}%
\usepackage{xcolor}%
\usepackage{textcomp}%
\usepackage{manyfoot}%
\usepackage{booktabs}%
\usepackage{algorithm}%
\usepackage{algorithmicx}%
\usepackage{algpseudocode}%
\usepackage{listings}%
\usepackage{siunitx}
\usepackage{comment}
\usepackage{hyperref}
\usepackage{pdflscape}
\usepackage{makecell}
\usepackage{multirow}
\usepackage{csquotes}
\usepackage{censor}
\usepackage{array}
\usepackage{caption}

\begin{document}

\title[Article Title]{Human Autonomy and Sense of Agency in Human-Robot Interaction: A Systematic Literature Review}

\author*[1]{\fnm{Felix} \sur{Glawe}}\email{glawe@comm.rwth-aachen.de}

\author[1]{\fnm{Tim} \sur{Schmeckel}}\email{schmeckel@comm.rwth-aachen.de}

\author[1]{\fnm{Philipp} \sur{Brauner}}\email{brauner@comm.rwth-aachen.de}

\author[1]{\fnm{Martina} \sur{Ziefle}}\email{ziefle@comm.rwth-aachen.de}

\affil[1]{\orgdiv{Chair for Communication Science}, \orgname{RWTH Aachen University}, \orgaddress{\street{Campus-Boulevard 57}, \city{Aachen}, \postcode{52074}, \country{Germany}}}

\abstract{Both human autonomy and agency are core human values and foundational to well-being. Their uses and definitions are intertwined and are central to current ethical and governance discourses on artificial intelligence and cyber-physical systems (e.g., the EU AI Act and IEEE Ethically Aligned Design). This review synthesises empirical studies on human autonomy and sense of agency in Human-Robot Interaction (HRI) to bridge the gap between existing design frameworks and regulatory demands on the one hand and the available empirical evidence on the other. Using the PRISMA workflow, we queried five databases and identified 22 empirical studies published between 2011 and early 2024. Across both concepts, separate methodological strands emerged: Self-determination theory-based measures of autonomy via psychometric scales and neuroscientific measures of agency, primarily intentional binding, with little integrative work bridging the two. A thematic synthesis identified five clusters of potentially influential factors: robot adaptivity, robot communication, robot anthropomorphism, exposure to robots, and individual differences. Evidence for the direction of effects varied across contexts, including education, health and care, industry, hospitality, delivery, and general social robotics. Mapping the findings onto the METUX framework showed that current work predominantly targets the interface and task spheres. The adoption, behaviour, life, and society spheres are largely unexamined, limiting understanding of longer-term, real-world impacts on well-being. We recommend integrating autonomy and agency paradigms in study designs for a more nuanced understanding of both concepts, standardising definitions and instruments, probing moderating user traits beyond demographics, and extending evaluations beyond short laboratory tasks into longitudinal, real-world settings. These steps could help advance autonomy-supportive robot design strategies that uphold ethical and psychological principles and foster user well-being in HRI.}


\maketitle

\section{Introduction}\label{Introduction}
Human-Robot Interaction (HRI) is a well-studied field. Robots are envisioned and studied for a broad range of applications, including robots designed to act as colleagues in assembly tasks, simple cleaning robots, and pet robots used in care settings \cite{weiss2021cobots, prassler2000short, vandemeulebroucke2018use}. Robots designed to work with or accompany humans must be adapted to humans' diverse needs to enable effective, human-centred interaction and collaboration \cite{kopp_2024, simoes2022designing}.

Numerous studies have examined how HRI and robots must be modelled and designed to foster user experience, acceptance of robots, trust in robots, and other context-specific goal variables such as increased learning or health benefits \cite{PRATI2021102072, lotz2019you, zhang2023sorry, lemaignan2016learning, jung2017exploration}. While acceptance of robots and the intended outcome of HRI are essential for the effective deployment of this technology, prioritising these aspects can neglect ethical concerns. 

In 2017, the Institute of Electrical and Electronics Engineers (IEEE), the largest technical professional association in the world, published "Ethically Aligned Design" to foster discourse and the development of autonomous technology aligned with specified ethical design principles \cite{IEEEEAD}. A central principle is the optimisation of technology for human well-being. The document argues that without well-being as a crucial aspect of technology design, even the best design intentions could have serious negative consequences. One crucial aspect mentioned to uphold human well-being is maintaining and encouraging human autonomy. Similarly, the EU's AI Act too, mentions human autonomy and agency as ethical principles of AI design \cite{europeanparliament2024}.  

\subsection{Human Autonomy}
The concept of human autonomy has a long history as a fundamental ethical principle, advanced by the highly influential groundwork of Immanuel Kant  \cite{schneewind1998invention, guyer2003kant}.He described human autonomy as freedom of action and choice, without being subject to the control of others or to one's own inclinations.  

Beyond its role as an ethical principle, autonomy is also identified as a basic psychological need (BPN) in the well-known self-determination theory (SDT) by Ryan and Deci \cite{ryan2000self}. In SDT, the psychological needs for autonomy, competence, and relatedness serve as prerequisites for intrinsic motivation, which in turn is an important factor for well-being. This relationship has been empirically verified across numerous contexts and cultures, demographic variables, and even differing levels of need desire and valuation \cite{RYAN2019111, chen2015basic}. Unlike desires, these psychological needs function as nutrients for optimal human functioning and have been conceptually and empirically shown to predict well-being universally \cite{ryan2022we, stanley2021meta, baard2004intrinsic}. In SDT, the need for autonomy is defined as the experience of willingness and volition, and the sense of integrity when actions, feelings, or thoughts are authentic and self-endorsed. When frustrated, it leads to experiences of conflict and pressure \cite{vansteenkiste2020basic}.

At this point, it is important to distinguish between machine and human autonomy. Machines can exhibit autonomy, while humans are in a state of autonomy but do not possess it as an attribute \cite{soma2022strengthening}. There exists even an automation conundrum: higher machine autonomy is associated with decreased situational awareness and reduced ability of human operators to take over control \cite{endsley2017here}. Balancing human and machine autonomy is therefore a crucial task. Moreover, human autonomy does not hinge on complete independence; it builds upon relationships and acknowledges that a human can never be fully independent of others \cite{peters2018designing}. For example, a well-trusted employer who recognises and shares employees' goals and values can foster human autonomy while remaining clear in their demands \cite{peters2018designing, heyns2018}. Furthermore, studies that conceptualise human autonomy within SDT provide no evidence of an oversaturation of autonomy satisfaction, given its consistently positive prediction of well-being metrics \cite{chen2015basic, peters2018designing}. This is corroborated by observations that satisfying human autonomy (and the other psychological needs) does not reduce the desire for even greater satisfaction \cite{sheldon2009psychological}. Instead, it is the frustration of human autonomy that predicts ill-being.  


Because of the pivotal role that human autonomy plays in predicting well-being, it serves as a foundational concept in human-computer interaction (HCI) \cite{bennett2023}. Accordingly, designing for human autonomy, and psychological needs in general, should be placed at the centre of robot design as well \cite{janssen2024psychological}. 

However, SDT and BPNs are rarely applied in the context of HRI, although evidence by Nikolova et al. suggests that robotisation at work in its current form might even lead to lowered human autonomy, relatedness, and meaningfulness \cite{janssen2024psychological, nikolova2024}. In contrast, theoretical considerations by Formosa highlight that robots could improve human autonomy by providing users with more valuable ends, improved autonomy-related competencies, and more authentic choices when used appropriately \cite{formosa2021robot}.  

Based on SDT, Peters et al. published the influential METUX framework in 2018 to guide HCI researchers and developers in designing technology for well-being based on the psychological needs of autonomy, relatedness, and competence \cite{peters2018designing}. In their framework, they posit six spheres of experience in which needs can be experienced and affected: adoption, interface, task, behaviour, life, and society. The adoption sphere refers to the experience of deciding to use a new technology; the interface sphere to the interaction with the technology via its interface; the task sphere to the experience of the task carried out with or by the technology; the behaviour sphere to the overarching behaviour supported or induced by the technology (e.g., cooking or sleep tracking); the life sphere to how the technology and the induced behaviour shape the user's life in general; and the society sphere to how society in general experiences the effects of the technology. While these spheres provide a way to separate potentially conflicting design effects on autonomy across different spheres and foster structured reasoning and evaluation, Peters et al. also emphasise that the spheres themselves are conceptual and their boundaries overlap.   

Using the METUX framework, Janssen and Schadenberg developed the first guidelines for well-being-oriented social robot design \cite{janssen2024psychological}. They based these guidelines on empirical studies in HRI concerned with related concepts such as empowerment, perceived freedom, and self-efficacy. However, they acknowledge that only a few articles have explicitly applied theories of well-being and motivation in HRI research. Although the guidelines are explicitly stated for social robots, we argue that they offer a valuable foundation for designing HRI across various domains: They present a testable, fixed set of rules, and there is evidence that even in highly industrial settings, people might relate to robots as social actors, potentially benefiting from design guidelines derived from social domains such as care or education \cite{sauppe2015social}.

For each sphere of experience (except the sphere of adoption and society), they define and provide examples of how the individual psychological needs of relatedness, autonomy, and competence can be satisfied in social robot design. Because this literature review is concerned with human autonomy, we summarise only the recommendations given to foster human autonomy: 

\begin{itemize}
    \item Interface Sphere: 
    \begin{itemize}
        \item Provide choice and options for interaction modes
        \item Apply non-controlling communication techniques
    \end{itemize}
    \item Task Sphere: 
    \begin{itemize} 
        \item Provide choices regarding tasks and task characteristics performed by the robot
        \item Acknowledge negative feelings towards a task
        \item Provide meaningful rationales for tasks
    \end{itemize}
    \item Behaviour Sphere: 
    \begin{itemize}
        \item Empower decisions about the type, difficulty and frequency of (users') behaviour
        \item Acknowledge negative feelings towards a behaviour
        \item Provide a meaningful rationale for behaviour
    \end{itemize}
    \item Life Sphere: 
    \begin{itemize}
        \item Prevent excessive engagement and overreliance on the robot
        \item Assess whether users still act in line with their own values.        
    \end{itemize}
\end{itemize}

Janssen and Schadenberg describe their work as a call to action to implement measures that foster psychological needs when designing robots and to use the framework as a reference point from which to start more structured and extensive research efforts \cite{janssen2024psychological}. 

\subsection{Sense of Agency}
The sense of agency, defined as the "feeling of control over actions and their consequences" \cite[p.1]{moore2016sense}, is often treated as synonymous with, or closely intertwined with, human autonomy, particularly in HCI \cite{bennett2023}.
Bennett et al. describe sense of agency and human autonomy in HCI as umbrella concepts: broad constructs that encompass a variety of theoretical perspectives and understandings, which may be related \cite{bennett2023}. They identify a shared focus on self-causality, perception of control, identity, and material independence \cite{cummins2014agency, bennett2023}. Additionally, both are regarded as intrinsically good and as contributing to physical and mental well-being \cite{bennett2023, moore2016sense}. 

While human autonomy as a psychological basic need is grounded in motivational psychology, the sense of agency is often approached from a neuroscientific perspective. Wegner and Wheatley proposed the Model of Conscious Will, which posits that unconscious causes of thought and action give rise to the thoughts and actions we experience \cite{wegner1999apparent}. Our sense of agency stems from consciously experienced thoughts and subsequent actions, yet the true causes of these actions are unconscious triggers that precede conscious intention \cite{moore2016sense}. 

In HCI, the sense of agency is acknowledged as an important factor in interface design \cite{moore2016sense}. Technology should allow users to feel agentic (i.e., in control). However, increasing automation in domains such as self-driving cars and flight-assistance systems can challenge this goal.

The fact that sense of agency and human autonomy function as vague umbrella concepts is not inherently problematic \cite{bennett2023}. When clearly articulated, such concepts can facilitate coordination and communication across disciplines. However, Bennett et al. identify a lack of integrative work needed to strengthen theoretical and empirical connections and clarify how these concepts relate to one another. For example, it remains an open question whether sense of agency constitutes a prerequisite for satisfying human autonomy---that is, whether the sense that we are the agents of our actions is necessary for perceiving those actions as aligned with our values and identity \cite{antusch2021intentional}.

\subsection{Study Aim}
With this study, we echo Janssen and Schadenberg's call for more empirical research on human autonomy and agency in HRI and aim to corroborate theoretical frameworks by synthesising existing empirical work. At present, design frameworks such as METUX and the derived guidelines for well-being-oriented social robots provide concrete recommendations for fostering human autonomy in HRI, yet these recommendations are grounded primarily in theoretical reasoning and evidence from adjacent fields rather than in direct empirical HRI research \cite{peters2018designing, janssen2024psychological}. Simultaneously, regulatory instruments including the EU's AI Act and the IEEE's Ethically Aligned Design require developers to attend to human autonomy and agency, but offer limited guidance on what empirically supports these goals in HRI \cite{europeanparliament2024, IEEEEAD}. As robots are increasingly deployed across domains, design and policy decisions concerning human autonomy and agency are already being made, yet without a synthesis of the available empirical evidence, it remains unclear to what extent existing frameworks are supported, where critical gaps persist, and which factors warrant further investigation. A systematic review that maps this evidence base is therefore needed to bridge the gap between theoretical and regulatory demands on the one hand and empirical knowledge on the other. Our review is guided by the following questions:
\begin{itemize}
    \item What domains, methods, conceptualisations, and operationalisations characterise the current empirical literature on human autonomy and agency in HRI?
    \item What factors have been examined as potentially influencing human autonomy and sense of agency in HRI, and what trends emerge across studies?
    \item How can the identified studies and findings be interpreted in light of the METUX framework and framework-derived design guidelines?
\end{itemize}

To address these questions, we identify and synthesise empirical research on how to maintain and foster human autonomy and sense of agency in HRI by applying the Preferred Reporting Items for Systematic Reviews and Meta-Analyses (PRISMA) guidelines \cite{page2021prisma}. Because human autonomy and sense of agency function as umbrella concepts in HCI, with potentially entangled or similar meanings, this review is designed to capture both concepts. In the results and discussion, we will highlight any meaningful differences and potential connections between those concepts. 

The results of this review can guide researchers, practitioners, and policymakers by highlighting a) research gaps for future studies and funding, b) aspects to prioritise in HRI design, and c) favourable measurement practices and instruments. 

The article is structured as follows: First, we describe our methodological approach based on the PRISMA workflow, including the utilised databases, deployed keywords, the process of article identification, characteristics of the final dataset, and the analytic procedures. Second, we present the results of the analysis. Finally, we discuss the results and the limitations of this study, interpret them in light of the METUX framework, and derive recommendations for subsequent research.

\section{Method}\label{Method}
In this section, we describe the process of this systematic literature review in accordance with the Preferred Reporting Items for Systematic Reviews and Meta-Analyses (PRISMA) guidelines \cite{page2021prisma}. We report the databases consulted as sources, the keywords used for initial identification, the screening and eligibility criteria, the procedures for assessing and classifying articles, and the content-coding approach.

\subsection{Databases}
To ensure comprehensive coverage of academic literature on human autonomy and agency in HRI, we searched multiple major databases for potentially relevant articles. The databases included Scopus, IEEE Xplore, Web of Science, ACM Digital Library, and ScienceDirect. Scopus, ScienceDirect, and Web of Science are broad, multidisciplinary databases of peer-reviewed academic output (approximately 23 to over 241 million indexed records), whereas IEEE Xplore and the ACM Digital Library focus on electrical engineering and computer science \cite{ACM2025, IEEE2025, elsevier2025, elsevier2025a, clarivate2025}. 

\subsection{Keywords}
The authors selected keywords based on their expertise in the domain. In addition, an iterative approach was used: relevant terms identified during data collection were added to retrieve additional articles in subsequent phases. This strategy aimed to ensure comprehensive coverage by incorporating newly acquired knowledge. Article identification proceeded in four phases. Keywords were queried in article titles, abstracts, and author keywords. In each phase, the search strings included \textit{robot and cobot} but varied the terms used to capture notions of human autonomy and agency. 

In the first phase, we used the following keywords: \textit{human autonomy, perceiving autonomy, perceived autonomy, perception of autonomy, user autonomy, personal autonomy, sense of autonomy, experiencing autonomy, experience of autonomy, supporting autonomy, support for autonomy, support of autonomy, preserving autonomy, preservation of autonomy, human agency, perceiving agency, perceived agency, perception of agency, user agency, personal agency, sense of agency, experiencing agency, experience of agency, supporting agency, support for agency, support of agency, preserving agency and preservation of agency}.

In the second phase, we used the keywords \textit{need for autonomy, need to feel autonomous and autonomy need} were used. 

The third phase used the keyword \textit{self-determination theory}. 

In the fourth and final phase, the keywords \textit{intrinsic motivation and basic psychological need} were used.

\subsection{Process of Article Identification}
\begin{figure}
\includegraphics[width = 230pt]{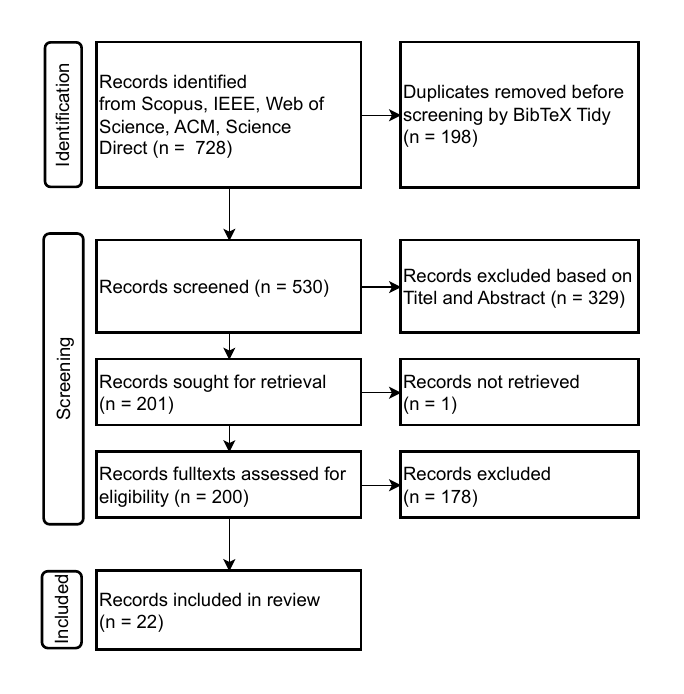}
\caption{Schematic depiction of the article identification and assessment process based on the PRISMA Flow diagram, resulting in a final set of n = 22 articles for analysis.} \label{fig1}
\end{figure}

In the following, we describe the process used to identify and assess articles for inclusion in this review. An overview of this process is depicted in Figure \ref{fig1}.
The four phases of article identification were conducted in a consistent manner by the first author, but in separate time frames. Phase one was conducted between 29.12.2023 and 08.03.2024, phase two between 08.04.2023 and 09.05.2024, phase three on 12.04.2024, and phase four on 27.05.2024. 
Because each database uses a different search syntax, the strings were adapted accordingly. For the exact search strings and settings used, please refer to the files uploaded to OSF.io.
Articles were recorded and saved as .bib files for each query if they met the following criteria:  
\begin{itemize}
    \item Article, conference paper, book chapter, or review
    \item Written in English
    \item Keywords present in the title, abstract, or keywords
    \item Not pertaining to human-autonomy teaming 
\end{itemize}

We excluded theses, preprints, technical reports, white papers, editorials, letters, and posters because they often lack formal peer review and therefore vary in quality. Although review articles typically do not present original empirical data, we included them in the search to mitigate the risk of missing a review with a similar scope to this work. No date range was applied in order to capture niche evidence and to trace historical trajectories and concept evolution.

In total, N = 728 articles were recorded. For each phase, records were merged and duplicates were automatically removed using the tool bibtex-tidy \cite{West2025}, resulting in 530 articles. Duplicates between phases were not removed at this stage because some phases were conducted in parallel. 
For the screening step, we established a fixed set of rules to select appropriate articles. These rules were derived from the goal of documenting the current empirical work on how human autonomy and agency are fostered in HRI. We explicitly focus on autonomous robots that do not require human control to capture cases in which robots might be perceived as colleagues or companions rather than as mere sophisticated tools:
\newpage
\begin{itemize}
    \item Accept:
    \begin{itemize}
        \item Addresses human autonomy or agency of a human in the context of HRI
        \item The robot or cobot refers to a singular entity capable of operating without being attached to a human body and without requiring control by a human operator
        \item Human autonomy or agency is measured after, during or for an imagined interaction with the robot or cobot
        \item The work is based on empirical data (e.g., interviews or video recordings, survey data, ...)
    \end{itemize}
    \item Reject: 
    \begin{itemize}
        \item Autonomy refers to the robot or the cobot 
        \item Addresses perceived autonomy or agency of the robot or cobot and not the human 
        \item Robots or cobots refer to digital assistants without a depictable physical body (virtual robots or fictive scenarios are accepted)
        \item The robot or cobot cannot potentially operate without being attached to a human's body or without being controlled by a human operator
        \item Human autonomy or agency is only used as a design guideline or is not measured after interaction
        \item The work is purely theoretical and does not rely on empirical data
   \end{itemize}
\end{itemize}

The .bib files for the 530 recorded articles were imported into Excel to facilitate assessment. Articles were then screened based on their title and abstract using the aforementioned rules and colour-coded accordingly (red = not eligible, green = further assessment, yellow = unsure + further assessment). This initial screening, including removal of any remaining duplicates, was conducted by the first author and yielded 201 articles for subsequent full-text assessment. The marked reduction was primarily due to the exclusion of articles addressing human-autonomy teaming, studies of robot agency or autonomy, and non-empirical work considering agency and human autonomy from ethical or theoretical perspectives. 
Full-text assessment was conducted independently by two coders (i.e., the first and second authors) following the same set of rules. All full texts were retrievable except for one. Each coder processed the database independently, after which classifications were compared. Discrepancies were documented and independently re-examined by both coders, who also recorded rationales for inclusion or exclusion of contested articles. These re-assessments were then compared and discussed. Following this process, both coders reached consensus on a final set of 22 articles, which was used for further analysis.

\subsection{Article Analysis}
The articles were analysed using qualitative content analysis, conducted independently by the two coders. Prior to coding, seven deductive categories were established: 

\begin{itemize}
\item Definition of human autonomy/agency
\item Operationalisation of human autonomy/agency
\item Type of robot 
\item Context of HRI
\item Method used
\item Independent variables
\item Results
\end{itemize}

The articles were imported into MAXQDA 2022 for coding. After individual coding, the coders compared and discussed the coded segments within each category. Discrepancies were resolved through discussion until consensus was reached. We refrained from calculating intercoder agreement statistics, given the factual nature of research articles. Intercoder agreement statistics calculate agreement and not correctness, and therefore misalign with the actual goal of data extraction. Subsequently, both coders inductively examined the content to identify common trends and patterns. These were also discussed, and the results are presented in the following chapter.

\section{Results}\label{Results}

\begin{sidewaystable*}[ph!]
\renewcommand\cellalign{tl}   
\setcellgapes{0pt}           
\makegapedcells
\setlength{\aboverulesep}{0pt}
\setlength{\belowrulesep}{0pt}
\caption{Articles and their descriptive features included in the literature review on human autonomy and agency in HRI.}\label{Articleoverview}%
\sisetup{
	detect-all,
    table-format = +.5
}
\begin{tabular}{l@{ } l l l l l l}
\toprule
Article & \makecell[l]{Year} & \makecell[l]{Context of HRI} & \makecell[l]{Type of Robot} & \makecell[l]{Concept \& Definition} & \makecell[l]{Operationalisation} & \makecell[l]{Method}\\
\midrule

\cite{donnermann2021} & 2021 & Education & Nao & \makecell[l]{Need for Autonomy\\ based on SDT (i.e. \cite{ryan2000self})} & Psychometric scale \cite{sheldon2008manipulating}& \makecell[l]{Experiment\\ with interaction}\\

\cite{henkemans2017} & 2017 & \makecell[l]{Health \& Care\\(Education)} & Nao & \makecell[l]{Need for Autonomy\\ based on SDT (i.e. \cite{ryan2000self})} & Psychometric scale \cite{la2000within}& \makecell[l]{Experiment\\ with interaction}\\

\cite{lee2023} & 2023 & Hospitality & Service Robots & \makecell[l]{Need for autonomy\\ based on SDT (i.e. \cite{ryan2000self})} & Psychometric scale \cite{nikou2017mobile, lee2018effect} & \makecell[l]{Survey\\ with video vignette}\\

\cite{lu2023} & 2023 & Education & Sanbot ELF & \makecell[l]{Need for autonomy\\ based on SDT (i.e. \cite{ryan2000self})} & Psychometric scale \cite{la2000within} & \makecell[l]{Experiment\\ with interaction}\\

\cite{moradbakhti2023} & 2023 & \makecell[l]{Health \& Care} & Pepper & \makecell[l]{Need for autonomy\\ based on SDT (i.e. \cite{ryan2000self})} & Psychometric scale \cite{moradbakhti2024development} & \makecell[l]{Survey\\ with video vignette}\\

\cite{ohshima2023} & 2023 & Social Robots & Pocketable Bones & \makecell[l]{Need for autonomy\\ based on SDT (i.e. \cite{ryan2000self})} &n.d.& \makecell[l]{Experiment and interview\\ with interaction}\\

\cite{sinai2022} & 2022 & Education & Nao & \makecell[l]{Need for autonomy\\ based on SDT (i.e. \cite{ryan2000self})}  & Psychometric scale \cite{mcauley1989psychometric} & \makecell[l]{Survey\\ with video vignette}\\

\cite{yang2023} & 2023 & Hospitality & Service Robots & \makecell[l]{Need for autonomy\\ based on SDT (i.e. \cite{ryan2000self})}  & Psychometric scale \cite{van2010capturing}& Survey\\

\cite{nikolova2024} & 2024 & Industry & Industrial Robots & \makecell[l]{Autonomy\\ based on \cite{hackman1976motivation}} & Psychometric scale \cite{nikolova2021perceived} & \makecell[l]{Secondary Database\\ Analysis}\\

\cite{koh2023} & 2023 & Health \& Care & Robotic Cats, PARO & \makecell[l]{Autonomy\\ based on \cite{beauchamp1994principles}} &n.d.& \makecell[l]{Interview\\ with video vignette}\\

\cite{coghlan2021} & 2021 & Health \& Care & ElliQ, Vector, Biscuit & \makecell[l]{Relational Autonomy\\ based on \cite{perkins2012relational}}& n.d. & \makecell[l]{Interview\\ with video vignette}\\

\cite{serholt2022} & 2022 & Education & Pepper & \makecell[l]{Autonomy\\ s.d.} & Psychometric scale n.d.& \makecell[l]{Experiment\\ with interaction}\\

\cite{barlas2019} & 2019 &n.d.& Nao & \makecell[l]{Sense of Agency\\ based on \cite{gallagher2000philosophical}} & \makecell[l]{IB paradigm and\\ Psychometric scale \cite{barlas2018action}} & \makecell[l]{Experiment\\ with interaction}\\

\cite{roselli2019} & 2019 & Social Robots & Cozmo & \makecell[l]{Sense of Agency\\ based on \cite{gallagher2000philosophical}} & IB paradigm & \makecell[l]{Experiment\\ with interaction}\\

\cite{sahai2023} & 2023 & Social Robots & Pepper & \makecell[l]{Sense of Agency\\ based on \cite{gallagher2000philosophical}}& IB paradigm & \makecell[l]{Experiment\\ with interaction}\\

\cite{ciardo2018} & 2018 &n.d.& Cozmo & \makecell[l]{Sense of Agency\\ based on \cite{frith2014action}}  & Psychometric scale n.d & \makecell[l]{Experiment\\ with interaction}\\

\cite{lombardi2023} & 2023 &n.d.& iCub & \makecell[l]{Sense of Agency\\ based on \cite{haggard2017sense}} & IB paradigm & \makecell[l]{Experiment\\ with interaction}\\
 
\cite{khalighinejad2016} & 2016 & n.d.& Robotic Hand & \makecell[l]{Sense of Agency\\ based on \cite{moretto2011experience}}& IB paradigm & \makecell[l]{Experiment\\ with interaction}\\

\cite{huete2011} & 2011 & Health \& Care & ASIBOT & \makecell[l]{Autonomy\\ n.d.} & Psychometric scale n.d. & User Test\\

\cite{kaur2023} & 2023 & Industry & n.d & \makecell[l]{Autonomy\\ n.d.} & Psychometric scale n.d. & Survey\\

\cite{swift2016} & 2016 & \makecell[l]{Health \& Care} & Nao & \makecell[l]{Autonomy\\ n.d.} & Psychometric scale \cite{grolnick1989parent} & \makecell[l]{Experiment\\ with interaction}\\

\cite{yu2023} & 2023 & Delivery & Starship & \makecell[l]{Sense of Agency\\ n.d.} & Psychometric scale \cite{tapal2017sense} & \makecell[l]{Experiment\\ with interaction}\\
\addlinespace[-13pt]
\end{tabular}
\footnotetext{\footnotesize{n.d.: not defined; s.d.: self defined; SDT: Self-determination Theory}}
\end{sidewaystable*} 

The final set of 22 articles identified through the PRISMA workflow was analysed; results are presented in this section (see Table \ref{Articleoverview}). Sections 3.1 through 3.6 address RQ1 by characterising the domains, methods, conceptualisations, and operationalisations of the included studies. Section 3.7 addresses RQ2 by synthesising the factors examined as potentially influencing human autonomy and sense of agency and the trends that emerge across studies. Because RQ3 requires interpretive mapping of findings onto the METUX framework, it is addressed in the Discussion (Section 4.3). We first describe the characteristics of the included articles. We then report the descriptive analysis of definitions and operationalisations of human autonomy and agency. Next, we outline the types of robots used and the contexts in which HRI occurred, followed by a review of the methodological approaches employed. Finally, we summarise the independent variables (where applicable) and the studies' main results.

\subsection{Article Characteristics}
\begin{figure}[h]
\centering
\includegraphics[width=210pt]{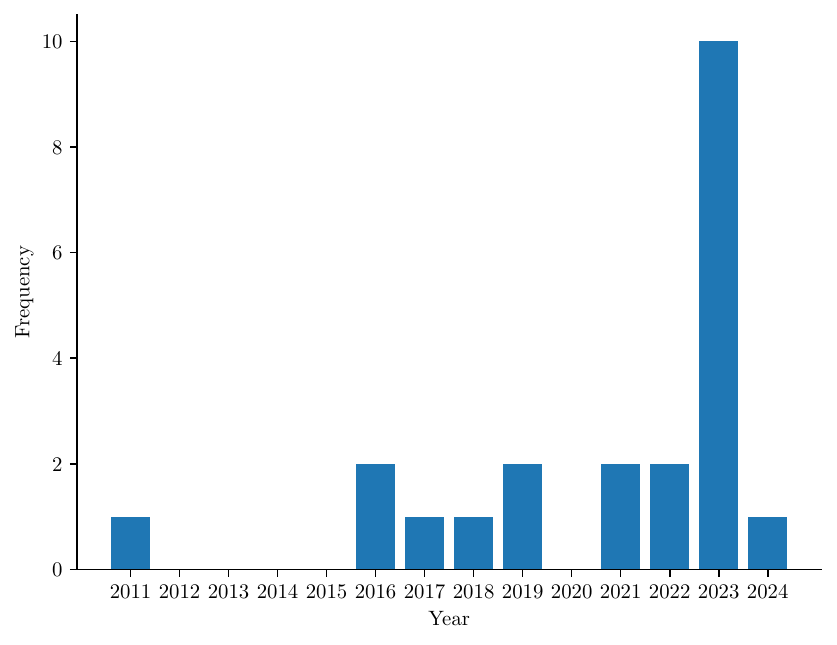}
\caption{Number of articles for each year on human autonomy and agency in HRI between 2011 and early 2024.} 
\label{fig2}
\end{figure}

The 22 articles were published between 2011 and 2024, with the largest share appearing in 2023 (10 articles; 45.5\%) (see Figure \ref{fig2}). Seven were conference papers (31.8\%), and fifteen were journal articles (68.2\%). Geographically, eleven studies were conducted in Europe (50\%), two in the USA (9.1\%), two in Australia (9.1\%), four in Asia (18.2\%), and three spanned multiple countries (13.6\%). 

\subsection{Contexts}
The studies, and thus the HRI examined, were situated within broader domains, including education, health and care, industry, hospitality, delivery, and general social robotics (see Table \ref{Articleoverview}). Four articles could not be assigned to a specific context. In one case, the health and care and educational contexts overlapped: a personal robot was designed and tested as part of an educational game on diabetes for affected children \cite{henkemans2017}. 

\subsection{Robot Types}
A variety of robot types were used or referenced in the analysed studies (see Table \ref{Articleoverview}). These included commercially available platforms such as the Nao robot by Aldebaran Robotics and Pepper by Aldebaran and SoftBank Robotics; self-developed models such as ASIBOT, an assistive robotic arm developed by the University of Getafe in Spain; a simple robotic hand; and robots with unspecified functions or general assistive purposes. The most frequently used and referenced platforms were Nao and Pepper.

\subsection{Definitions of Human Autonomy and Agency}
Two predominant definitions emerged (see Table \ref{Articleoverview}). Human autonomy was most often defined within self-determination theory as a basic psychological need—an experience of willingness, volition, and integrity when actions, feelings, or thoughts are authentic and self-endorsed \cite{vansteenkiste2020basic, ryan2000self}.
Sense of agency was predominantly defined as the feeling of control over one's own actions and their consequences \cite{haggard2017sense, frith2014action, moretto2011experience}. In four cases, alternative definitions of human autonomy were used. One employed the definition by Beauchamp and Childress, which emphasises free choice \cite{beauchamp1994principles}. The remaining three adopted context-specific definitions: in care, relational human autonomy was used, defined as the ability to make decisions with input and support from external sources \cite{coghlan2021}; in industrial settings, Nikolova et al. translated human autonomy as the freedom to judge and determine how to work; and in education, Serholt et al. defined human autonomy as the tutor's ability to fulfil their role without external help \cite{nikolova2024, serholt2022}. Four articles provided no definitions of either human autonomy or agency (see Table \ref{Articleoverview}).

\subsection{Applied Methods}
The primary methodology was experimentation involving on-site interaction between participants and robots (see Table \ref{Articleoverview}). Other studies deployed surveys, either accompanied by video vignettes that provided a scenario for evaluation, or targeting participants’ past and ongoing experiences with robots in their daily or work lives. One study conducted a user test without a formal experimental framework, and another combined survey data with secondary data on industrial robotisation. Three studies utilised interviews; two of these incorporated video vignettes. In one case, interviews were conducted post-HRI within an experimental framework.

\subsection{Operationalisation of Human Autonomy and Agency}
Operationalisation primarily relied on psychometric scales or the intentional-binding paradigm, with one study employing both approaches concurrently (see table \ref{Articleoverview}). The intentional binding effect refers to the perceived compression of the interval between a voluntary action and its external consequence and is commonly used as a measure of sense of agency \cite{MOORE2012546}. Five studies assessed sense of agency using the intentional binding paradigm, albeit with varying procedures. In three, participants explicitly estimated the elapsed time between their action and its outcome \cite{barlas2019, lombardi2023, sahai2023}. In the other two, the interval was inferred from the difference between the participant-indicated and actual positions of a clock hand \cite{khalighinejad2016, roselli2019}. In two studies, the estimated interval served as a direct proxy for sense of agency \cite{lombardi2023, sahai2023}. In the remaining studies, the interval (or clock-position discrepancy) was baseline-corrected by subtracting values obtained in a condition without human action, in which actions were simulated. Most tasks paired a specific action (e.g., a timed keypress) with a simple outcome (e.g., a tone). Depending on the research aim, the task was performed with, alongside, or under the guidance of a robot. In Lombardi et al., the action was directed at the robot (the participant looked at the robot), and the outcome was the robot’s subsequent gaze direction or facial expression \cite{lombardi2023}.  

A variety of instruments provided explicit measures via psychometric scales (see Table \ref{Articleoverview}). Most studies used or adapted existing scales; however, only two articles employed the same instrument by La Guardia et al. \cite{la2000within}. Notably, the four studies that did not use established instruments relied on single items, whereas the remaining scales used at least three items. Recurring item themes included perceived control, (in-)dependence, choice, being oneself, free will, alignment of actions with the self, and felt ownership of actions. Among the three articles that assessed sense of agency with psychometric scales, two used a single item focused solely on perceived control \cite{barlas2019, ciardo2018}, whereas Yu et al. employed a multi-item scale capturing perceived control, free will, responsibility for actions, intentionality, and planned behaviour \cite{yu2023}. While perceived control and free will overlap across operationalisations of sense of agency and human autonomy, elements such as the freedom to be oneself and to act in line with one’s true self appeared only in operationalisations of human autonomy \cite{yang2023, henkemans2017}.

Three studies conducted qualitative interviews but did not specify the criteria used to classify participants’ statements as pertaining to human autonomy or sense of agency (see Table \ref{Articleoverview}). 

\subsection{Human Autonomy and Agency Influencing Factors}
\begin{figure*}[h]
\centering
\includegraphics[width=300pt]{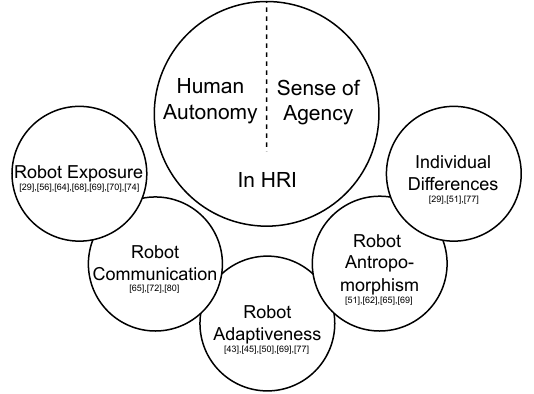}
\captionsetup{justification=centering}
\caption{Identified clusters of potential human autonomy and agency influencing factors in HRI.} \label{fig3}
\end{figure*}

Based on the factors examined across the included studies, we identified five clusters of variables potentially related to human autonomy or sense of agency (see Figure \ref{fig3}). These clusters are broad but not exhaustive. Accordingly, every article was assigned to at least one cluster. Studies that measured human autonomy or sense of agency following interaction but did not isolate specific influencing factors were placed in the 'Other' category. 

\subsubsection{Robot Exposure}
Seven studies examined the effects of interacting with, observing, or merely co-existing with a robot on sense of agency or human autonomy \cite{ciardo2018, khalighinejad2016, roselli2019, serholt2022, yang2023, nikolova2024, sinai2022}. Observing a robotic hand perform an action similar to one’s own increased sense of agency, mirroring the effect of observing a human hand \cite{khalighinejad2016}. In a survey conducted in the hospitality and tourism industry in China, employees in roles involving service robots reported higher human autonomy \cite{yang2023}. By contrast, robotisation negatively affected human autonomy when applied to repetitive, monotonous tasks; paired with social tasks or computer-based work, robotisation showed positive effects \cite{nikolova2024}. In a tutoring context, children experienced lower human autonomy when tutoring a robot compared with tutoring a human \cite{serholt2022}. Two independent experiments reported a reduced sense of agency when performing a task with a robot versus alone \cite{ciardo2018, roselli2019}. Finally, deploying a social robot as an online learning companion with basic psychological need-supportive behaviours did not affect human autonomy \cite{sinai2022}.  

\subsubsection{Robot Communication}
Robot communication was investigated in three articles, all referring to sense of agency \cite{lombardi2023, barlas2019, yu2023}. In an experiment on movement-path communication by delivery robots in urban spaces, displaying a robot’s predicted movement path increased sense of agency, whereas displaying the predicted path of the approaching human decreased it \cite{yu2023}. In two experiments, positive or negative facial expression of a robot was examined, but no effect on sense of agency was found \cite{lombardi2023, barlas2019}. However, Lombardi et al. showed that gaze direction mattered: sense of agency was significantly lower for an averted gaze than for direct gaze (eye contact) \cite{lombardi2023}.

\subsubsection{Robot Adaptiveness}
Robot adaptiveness, adjusting behaviour to users’ actual or presumed needs, was examined in five studies, all based on the concept of human autonomy \cite{donnermann2021, henkemans2017, kaur2023, lu2023, sinai2022}. An adaptive tutoring robot offering explanatory feedback on incorrect answers yielded only descriptively higher human autonomy than a non-adaptive tutor \cite{donnermann2021}. A personal robot that exhibited human-autonomy-satisfying behaviours (alongside competence- and relatedness-satisfying behaviours), such as offering choices to continue or pause, acknowledging mood, and integrating personally relevant scenarios, significantly increased human autonomy in a diabetes-related educational game \cite{henkemans2017}. In work settings, assistance provided on an as-needed basis was associated with higher human autonomy than no assistance or continuous assistance \cite{kaur2023}. By contrast, highly interactive support in adult education (answering domain-specific questions, allowing users to skip questions, and integrating videos/tutorials) did not increase human autonomy \cite{lu2023}. Offering choice alone also had no discernible impact on perceptions of a learning companion robot \cite{sinai2022}.

\subsubsection{Robot Anthropomorphism}
Four studies investigated anthropomorphism, the attribution of human characteristics to robots, in relation to human autonomy and sense of agency \cite{coghlan2021, moradbakhti2023, sahai2023, barlas2019}.  Saha{\"\i} reported a higher sense of agency when interacting with a human or a human-like robot compared with a servomotor \cite{sahai2023}. Perceived human-likeness of a robot in a video vignette correlated moderately with human autonomy \cite{moradbakhti2023}. In contrast, sense of agency did not differ when task instructions were given by a human versus a robot and was highest when no instructions were given at all \cite{barlas2019}. The perceived autonomy of the robot itself did not affect the sense of agency. In a qualitative study of companion robots for older adults, participants indicated that a humanoid appearance could either increase or decrease human autonomy depending on individual preferences and whether the robot was perceived as paternalistic \cite{coghlan2021}.  

\subsubsection{Individual Differences}
Three studies explicitly examined individual differences \cite{kaur2023, moradbakhti2023, nikolova2024}. Among people with prior experience working with cobots, Kaur et al. found no overall differences in human autonomy by age or gender; however, as-needed assistance increased human autonomy among female and older respondents, whereas no-assistance settings elicited more human autonomy among younger respondents \cite{kaur2023}. Moradbakhti et al. observed that, for male participants, attributing masculinity to a care robot was positively associated with higher human autonomy, with a weaker effect for women; moreover, men’s propensity for stereotyping was inversely associated with human autonomy in HRI \cite{moradbakhti2023}. In a study of general robotisation effects, no gender differences emerged; instead, higher education moderated the negative impact of robotisation on human autonomy \cite{nikolova2024}. 

\subsubsection{Other}
Five studies did not fit the identified clusters \cite{lee2023, huete2011, koh2023, ohshima2023, swift2016}. These works observed human autonomy but did not link it to specific factors or compare outcomes with non-robot control groups. In a study conducted in a hospitality context, participants imagining interactions with a service robot reported satisfactory human autonomy \cite{lee2023}. Users of an in-home assistive robot for disabled or older adults also reported experiencing human autonomy \cite{huete2011}. Deploying a socially assistive robot as an exercise buddy for overweight adolescents yielded a descriptive increase in human autonomy from pre- to post-intervention \cite{swift2016}. In a qualitative study, participants reported feelings of autonomy when walking with a self-augmentation robot that synchronised its gaze with the user during side-by-side communication and autonomously turned its gaze to show interest upon detecting a person ahead \cite{ohshima2023}. Finally, in dementia care, pet robots appeared to induce resident happiness, which in turn could increase staff human autonomy by facilitating connection with residents \cite{koh2023}.   

\section{Discussion}\label{Discussion}
Human autonomy is a basic psychological need that, when satisfied, fosters intrinsic motivation and overall well-being. Human autonomy and agency are also conceptualised as fundamental ethical principles and are embedded in regulatory frameworks such as the EU's AI Act and the IEEE's Ethically Aligned Design \cite{europeanparliament2024, IEEEEAD}. At the same time, robots are assuming an increasingly pivotal role across many domains and are expected to become even more central. It is therefore surprising that relatively few empirical studies were identified, especially compared with the extensive work on other human factors in HRI (e.g., trust, acceptance) \cite{naneva2020systematic}. As this review shows, publications have increased in recent years—particularly in 2023, suggesting a gradual shift in research attention toward human autonomy and sense of agency in HRI. Most studies were conducted in Europe, particularly within the European Union, potentially reflecting the EU's push for human-centred innovation \cite{EU2024, europeancommission2025, europeancommission2025a}. A range of empirical methods was used, predominantly surveys and experiments, with fewer interviews—a noteworthy pattern for a fragmented field in which one might expect more exploratory qualitative work. Across contexts (education, health and care, industry, hospitality), studies were relatively evenly distributed, and we observed no consistent relationships between domains and outcomes. The frequent use of commercially available robots such as Pepper, Nao, and Coszmo likely reflects their accessibility and cost-effectiveness. While converging on similar platforms can facilitate comparability, it may also constrain methodological flexibility. 

\subsection{Human Autonomy and Sense of Agency: Conceptual and Methodological Distinctions}
The definitions used for human autonomy and sense of agency reflect two largely separate research strands: one grounded in human autonomy (or the need for autonomy) as a basic psychological need rooted in SDT within motivational psychology, and the other grounded in the neuroscientific concept of sense of agency. We did not identify work that integrates these perspectives. Regarding operationalisation, sense of agency was predominantly measured implicitly via the intentional binding paradigm, whereas studies on human autonomy relied exclusively on psychometric scales. In the few cases where sense of agency was assessed with a psychometric scale, we observed both overlap and divergence relative to human autonomy measures. Overall, the operationalisations highlighted differences more than similarities. Nevertheless, perceived control appears to be a common denominator, defining for sense of agency and, at least instrumentally, an important aspect of human autonomy as well \cite{moore2016sense, bennett2023}. This overlap reinforces the question raised in the Introduction of whether sense of agency may constitute a prerequisite for human autonomy, a relationship that remains to be empirically tested \cite{antusch2021intentional}. Given these methodological and theoretical divergences, this section contributes by documenting the distinct research strands and identifying perceived control as a shared element. We treat human autonomy and sense of agency separately in the sections that follow, not because they necessarily represent distinct constructs, but because the reviewed evidence does not yet permit conclusions about their equivalence or hierarchy.
  
\subsection{Human Autonomy and Agency Influencing Factors}
Although research on the sense of agency and human autonomy in HRI remains limited, five thematic clusters of examined factors can be identified: robot adaptiveness, robot communication, robot exposure, robot anthropomorphism, and individual differences. No cluster showed a clear trend regarding whether specific factors positively or negatively affect human autonomy or sense of agency. Nevertheless, these clusters provide an initial set of factors worthy of deeper investigation. 

\textbf{Robot Exposure:} Many studies measured the effects of interacting with, working alongside, or observing a robot on human autonomy or sense of agency. As expected—given the absence of explanatory variables in some studies and the heterogeneity of scenarios (e.g., observing a robotic hand vs. robotisation in industry) \cite{nikolova2024, khalighinejad2016}, the findings were diverse. Consequently, no general conclusion can be drawn about the direction of robots’ influence on human autonomy or sense of agency per se. Still, it shows that exposure to robots can alter perceptions of autonomy and sense of agency. The direction of this effect likely depends on how the robot is deployed and used—for example, whether users gain meaningful choices \cite{formosa2021robot}. Moreover, research on acceptance and trust indicates that individual differences (e.g., attitudes toward and experience with robots) shape how HRI is perceived, and a comparable influence is conceivable for human autonomy and sense of agency \cite{bloch2023comparison, nikolovska2024user}.   

\textbf{Robot Communication:} Janssen and Schadenberg’s guidelines for well-being-oriented social robots underscore the importance of communication and its potential impact on human autonomy across all METUX spheres \cite{janssen2024psychological}. They emphasise non-controlling styles, acknowledging human feelings, and providing rationales for actions and requested behaviours. Notably, we found no study directly examining robot communication effects on human autonomy, only on sense of agency. Yu et al. demonstrated how controlling communication can be as subtle as visualising an anticipated human movement path \cite{yu2023}. Although participants were not instructed to follow the path, they felt pressured to comply with the robot’s prediction, lowering their sense of agency. 
Gaze, an essential component of human communication, also influenced sense of agency \cite{lombardi2023}. Eye contact is hypothesised to trigger self-awareness and thus elevate sense of agency via self-referential processing \cite{BALTAZAR2014120, ULLOA2019102794, CONTY2016184}. Gaze further signals intentions, supports turn-taking, and can help build a social bond between interaction partners \cite{yoshikawa2006responsive, ccakir2023reviewing}. Such a bond may instil relatedness, another psychological need that can, in turn, bolster human autonomy \cite{ryan2000self}. Interestingly, a robot’s facial expression (positive or negative) did not affect sense of agency \cite{lombardi2023, barlas2019}, which contrasts with theories suggesting people claim responsibility for positive outcomes (i.e. a positive expression) more than negative ones to protect self-esteem \cite{greenberg1992depression}. Because the valence of the robot’s expression was randomised, participants could not predict outcomes. Consistently, Yoshie and Haggard found that when emotional outcomes are predictable, negative outcomes reduce sense of agency, whereas randomised outcomes show no such effect \cite{YOSHIE20132028}. 

\textbf{Robot Anthropomorphism:} Anthropomorphism is a well-known design factor, including the often undesirable uncanny valley effect \cite{Mori201298}. Extensive research has shown that anthropomorphism can shape perception and acceptance \cite{siegel2009persuasive, Roesler2021}. In this review, no clear effect emerged for either sense of agency or human autonomy, yet several mechanisms are plausible. At the interface level, greater anthropomorphism may enhance intuitiveness and free cognitive resources, fostering feelings of independence and choice. Conversely, anthropomorphic robots may be perceived as threats to one’s abilities or as paternalistic \cite{coghlan2021}. Human-like features can also blur the line between robots as tools and as living beings, potentially encouraging deception or undesirable user behaviour, such as diverging from one’s own values \cite{Hegel2008574, DANG2024123683, bao2023mitigating}. Decisions about the level and specific features of anthropomorphism, therefore, warrant careful consideration. 

\textbf{Robot Adaptiveness:} Robots’ adaptiveness to users’ needs can enhance human autonomy, but only when guided by effective interventions, sound design choices, and sensitivity to individual differences. The mixed results in this review illustrate the challenge. Henkemans et al. showed that adopting recommendations from related domains (e.g., educational practice) can be a useful starting point for designing robots that foster human autonomy \cite{henkemans2017}. As noted by Formosa, the nature of the choice offered (meaningful vs. unimportant) is crucial \cite{formosa2021robot}. Simply adding options does not necessarily increase human autonomy \cite{sinai2022}. Although we did not identify HRI studies examining adaptiveness effects on sense of agency, HCI findings indicate that technological or algorithmic assistance does not monotonically increase sense of agency \cite{coyle2012did, ueda2021influence}. Instead, a threshold or tipping point appears beyond which assistance becomes counterproductive. Similar HRI experiments could help calibrate adaptiveness so that assistance remains below this tipping point.. 

\textbf{Individual Differences:} Demographic variables such as age, gender, and education play a substantial role in technology design and acceptance, and this review suggests the same may hold for human autonomy in HRI \cite{szalma2009individual, venkatesh2012consumer}. However, no clear trends emerged for age or gender, and only one study identified a moderating role of education. The three studies addressing individual differences focused on human autonomy, though evidence suggests that age may also relate to variations in sense of agency \cite{mariano2024aging}. Nevertheless, it is important to note that it is the demographic variables underlying values, beliefs or changes in cognitive abilities that are the reason for their influence \cite{salthouse2003needs, czaja2006factors}. Thus, examining demographics alone is insufficient to explain their potential influence on sense of agency and human autonomy.  

To summarise, interacting with, working alongside, or even merely observing robots can affect human autonomy and sense of agency; the extent and direction of these effects depend on how robots are implemented and deployed. Subtle controlling cues in communication can undermine agency, whereas establishing gaze contact may offer a straightforward means to foster it. Anthropomorphic features should be introduced cautiously and tested thoroughly. While adaptiveness is generally beneficial, assistance should not be treated as a “more is always better” feature. Finally, a nuanced understanding of prospective users and the individual differences that shape their experiences is essential.

\subsection{Links to METUX}
The METUX framework by Peters et al. delineates six spheres of experience through which technology can affect the psychological needs of autonomy, competence, and relatedness, as established in SDT. Building on this framework, Janssen and Schadenberg proposed design guidelines for social robots. We connect the results and articles identified in this review to METUX in two ways: first, by assessing how human autonomy and sense of agency were measured and to which interaction sphere they relate; second, by examining HRI designs and implemented factors and linking them to the guidelines and their respective spheres.

First, none of the included studies explicitly grounded their work in METUX, employed its sphere-specific instruments, or used the guidelines to design interactions. Nevertheless, most measures used map to the interface or task spheres. Only a few items unambiguously align with a single sphere (e.g., “I am able to choose or change the order of tasks” \cite{nikolova2024}). In most psychometric scales, the contextual detail was insufficient to distinguish whether items referred to the robot’s interface or the task performed with the robot (e.g., “The service robot gives me various options and choices.” \cite{lee2023}). This lack of differentiation complicates interpretation because enhancing human autonomy in one sphere may not translate to, and can even undermine, autonomy in another \cite{peters2018designing}. Sense of agency operationalised via the intentional binding paradigm presents an instructive case: intentional binding captures the immediate, pre-reflective coupling between an action and its effect, primarily reflecting interface-level interaction mechanics. It may also secondarily map to the task sphere when interpreted as felt control over a robot-supported task. Crucially, it remains to be tested whether sense of agency at the interface or task level relates to the satisfaction of human autonomy. If so, intentional binding could be a valuable complement to instruments measuring human autonomy in these spheres. Regarding the adoption sphere, only Coghlan et al. included an open-ended question capable of revealing autonomy-related adoption patterns (“Would you want to use them and why? If not, why not?” \cite[p.5]{coghlan2021}), yet their results did not indicate whether adoption was controlled or autonomous \cite{ryan2018, peters2018designing}. Adoption in HRI currently appears more frequently framed through the Technology Acceptance Model (TAM) or Unified Theory of Acceptance and Use of Technology (UTAUT) than through human autonomy or sense of agency \cite{he2022technology, concetta2025toward, DEGRAAF20131476}. At the other end of METUX, the behaviour and life spheres concern sustained, real-world behaviour change and broader well-being. HRI evaluations remain predominantly short, lab-based, and task-focused, mapping neatly onto METUX’s interface and task spheres but not to the longer time horizons required for behaviour and life. Accordingly, only the interviews in Koh et al. and Coghlan et al. offer insights aligned with these spheres: Koh et al. reviewed staff reflections on longer-term, real-world interactions in dementia care, and Coghlan et al. elicited prospective reflections on anticipated long-term effects \cite{koh2023, coghlan2021}. 

Comparing the implemented HRI factors in the articles with the guidelines revealed overlaps and differences. For the interface sphere, the findings support the proposition that non-controlling communication and offering options for interaction modes could contribute to human autonomy at the interface level \cite{huete2011, yu2023}. In addition, studies on gaze direction, robot gender, and perceived human-likeness suggest further interface-level levers to foster human autonomy and sense of agency beyond the current guidelines \cite{moradbakhti2023, sahai2023, lombardi2023}. At the task level, Henkemans et al. illustrate the effectiveness of guideline-aligned design, i.e. providing meaningful choices, acknowledging moods and feelings, and offering rationales grounded in personal scenarios \cite{henkemans2017}. Other studies similarly underscore the importance of choice over robot-performed tasks and task characteristics \cite{huete2011, kaur2023}. Findings by Nikolova et al. further indicate that task type and individual differences can moderate outcomes, underscoring that “one-size-fits-all” solutions are unlikely \cite{nikolova2024}. Unlike at the interface level, we did not identify task-level aspects that clearly challenge or expand the proposed guidelines. With respect to the behaviour sphere, Swift-Spong et al. deliberately integrated affective, positive messaging from the robot to promote a target behaviour (increased exercise) \cite{swift2016}. While Janssen and Schadenberg assign positive feedback to the interface sphere as competence support, Swift-Spong et al.’s rationale suggests such feedback may also be beneficial at a behavioural level. Finally, qualitative interviews in Coghlan et al. and Koh et al. highlight potential threats to autonomy at the life sphere, over-attachment disrupting routines and assistance contributing to cognitive or physical decline and increased dependence, both of which are acknowledged in the guidelines \cite{coghlan2021, koh2023}.  

In summary, none of the articles explicitly applied METUX; most focused on interface and task spheres, often without a clear distinction, complicating interpretation. Sense of agency, frequently assessed via intentional binding, remains insufficiently linked to broader human autonomy satisfaction and warrants further investigation. Adoption was rarely examined through an autonomy lens, and behaviour and life spheres were addressed only sporadically via qualitative work, revealing a gap in long-term, real-world evaluations. When comparing implemented HRI factors with existing guidelines, we found support for established recommendations at the interface and task levels, and identified additional aspects (e.g., gaze direction, robot anthropomorphism) that could enrich guideline development. At the behavioural level, positive feedback may promote desired outcomes beyond competence support. Identified threats at the life sphere, such as over-attachment or increased dependence, align with and further motivate the guideline cautions. 

\subsection{Limitations}
This review has several limitations. Despite an iterative keyword strategy intended to ensure comprehensive coverage, two relevant articles were identified post hoc via colleagues and were not captured in the initial search. Turja et al. examined differences in satisfaction of the need for human autonomy between robotised and non-robotised workplaces in Finland, finding that psychological needs, including autonomy, were more frequently satisfied in non-robotised workplaces \cite{turja2022basic}. Dammers et al. reported a user study testing the impact of different collaboration types in HRI on perceived human autonomy; they found no significant differences across collaboration types, although the highest level of collaboration yielded the descriptively highest human autonomy \cite{dammers2022usability}. The first article was missed because it used the term “basic needs” rather than “basic psychological needs,” and the second was not indexed in any of the five databases searched. This underscores the importance of using widely adopted terms while accounting for simple lexical variants and of ensuring that articles are indexed in major databases.

Although the authors, as two independent coders, conducted the full-text assessment and final coding, the initial identification and first screening were not independently duplicated, which may have introduced selection bias. Moreover, while discrepancies were reconciled through discussion, inter-rater reliability statistics were not computed. A formal quality or bias assessment was also not performed, which may have led to inclusion of studies with quality issues (e.g., single-item measures \cite{huete2011}). However, our objective was to foreground thematic patterns, lay a foundation for a joint research area on human autonomy and agency, and acknowledge heterogeneity rather than impose narrow quantitative quality judgments.

Our review drew on five large scientific databases; nevertheless, constraining the source pool may have excluded relevant work elsewhere. 

Finally, we did not conduct a meta-analysis of quantitative results. Although meta-analysis can strengthen inferences, it was not appropriate here due to pronounced conceptual and methodological heterogeneity across the reviewed articles. 

\subsection{Research Opportunities and Recommendations}
While human factors such as acceptance and trust in HRI have been extensively studied, comparable work on human autonomy and sense of agency remains limited. Given the importance of human autonomy, there is a clear need to determine how robots should be designed, adapted, and introduced to support it. This review represents an initial effort to collate and cluster empirical research on the topic. Although we identified clusters of potential influencing factors, the field is still emerging. We therefore outline several research opportunities and recommendations.  

First, investigate whether and how sense of agency and human autonomy relate to one another. Empirical studies that integrate both concepts could help establish whether a systematic connection exists or contribute to a more nuanced understanding of their distinctiveness.

Second, adopt established definitions and instruments to build a more comparable evidence base. Standardisation will increase the visibility of relevant research and facilitate future meta-analyses. 

Third, develop a fine-grained understanding of the factors that contribute to human autonomy and sense of agency in HRI. Empirical work should leverage frameworks such as METUX and the derived guidelines, and draw on related findings (e.g., autonomy-supporting behaviour in work contexts; acceptance and trust research) \cite{janssen2024psychological,peters2018designing,naneva2020systematic}. In addition, exploratory and qualitative approaches are valuable for uncovering candidate factors and mechanisms, reducing the likelihood of “one-shot” studies \cite{janssen2024psychological}.

Finally, investigate individual differences beyond demographics, such as the Big Five and causality orientation, to understand how these traits relate to variations in human autonomy and sense of agency and to advance truly human-centred HRI \cite{anglim2020predicting, koestner2023causality}. 

\section{Conclusion}\label{Conclusion}
The exploration of human autonomy and sense of agency in HRI is increasingly significant, as robots are poised to become indispensable across domains. This literature review underscores the urgent need to design and introduce robots in ways that preserve and promote people’s autonomy and sense of agency, thereby supporting overall well-being.

The present findings indicate that empirical research in this area remains nascent, with two primary strands: one rooted in self-determination theory, examining human autonomy (or the need for autonomy) via psychometric measures; and the other grounded in the neuroscientific notion of sense of agency, often employing the intentional binding paradigm. Integrative work bridging these strands is currently absent. Evidence suggests that factors such as robot adaptiveness, communication style, anthropomorphism, and individual user differences shape how human autonomy and sense of agency are perceived and experienced in HRI settings. However, the literature is relatively fragmented, with substantial variation in contexts, study designs, and instruments. Mapping our findings to the METUX framework indicates that current articles mainly address interface and task spheres. Comparing results with METUX and SDT-derived social robot design guidelines support elements such as non-controlling communication and meaningful choices, while highlighting opportunities for extension regarding anthropomorphism and positive feedback.

Building on these observations, future research should investigate whether and how sense of agency and human autonomy relate to one another. Furthermore, definitions and instruments should be standardised to enhance the reliability of cross-study comparisons. By conducting further exploratory research, integrating factors identified in contexts outside HRI, and applying existing theoretical frameworks (e.g., METUX and the design guidelines), the set of influential factors can be broadened and leveraged in forthcoming studies. 

Technology should be designed and deployed to foster human well-being; without this overarching goal, technological progress loses its purpose. Robots are no exception, and, given their envisioned roles as companions or colleagues, occupy a special position. Consequently, they are poised to intersect with aspects of human life that have historically been exclusive to human interaction.

\backmatter

\bmhead{Data availability}

The data and category system are publicly accessible at the Open Science Framework repository \url{https://osf.io/r7p28/?view_only=d569ab57f4164903a8150a3f67b44363}. 

\bmhead{Funding}

Funded by the Deutsche Forschungsgemeinschaft (DFG, German Research Foundation) under Germany’s Excellence Strategy– EXC-2023 Internet of Production – 390621612.

\section*{Declarations}

\bmhead{Conflict of interests}
The authors have no competing interests to declare that are relevant to the content of this article.

\bmhead{Ethics Declaration}
Not applicable.

\bibliography{References}

@article{page2021prisma,
	title        = {The {{PRISMA}} 2020 statement: an updated guideline for reporting systematic reviews},
	author       = {Page, Matthew J and McKenzie, Joanne E and Bossuyt, Patrick M and Boutron, Isabelle and Hoffmann, Tammy C and Mulrow, Cynthia D and Shamseer, Larissa and Tetzlaff, Jennifer M and Akl, Elie A and Brennan, Sue E and others},
	year         = 2021,
	journal      = {BMJ},
	publisher    = {British Medical Journal Publishing Group},
	volume       = 372,
	number       = 71,
	pages        = {1--9},
	doi          = {10.1136/bmj.n71}
}

@misc{West2025,
	title        = {bibtex-tidy},
	author       = {West, Peter},
	year         = 2025,
	publisher    = {GitHub},
	url          = {https://flamingtempura.github.io/bibtex-tidy}
}

@article{vansteenkiste2020basic,
	title        = {Basic psychological need theory: Advancements, critical themes, and future directions},
	author       = {Vansteenkiste, Maarten and Ryan, Richard M and Soenens, Bart},
	year         = 2020,
	journal      = {Motivation and emotion},
	publisher    = {Springer},
	volume       = 44,
	number       = 1,
	pages        = {1--31},
	doi          = {10.1007/s11031-019-09818-1}
}

@article{weiss2021cobots,
	title        = {Cobots in industry 4.0: A roadmap for future practice studies on human--robot collaboration},
	author       = {Weiss, Astrid and Wortmeier, Ann-Kathrin and Kubicek, Bettina},
	year         = 2021,
	journal      = {IEEE Transactions on Human-Machine Systems},
	publisher    = {IEEE},
	volume       = 51,
	number       = 4,
	pages        = {335--345},
	doi          = {10.1109/THMS.2021.3092684}
}

@article{prassler2000short,
	title        = {A short history of cleaning robots},
	author       = {Prassler, Erwin and Ritter, Arno and Schaeffer, Christoph and Fiorini, Paolo},
	year         = 2000,
	journal      = {Autonomous Robots},
	publisher    = {Springer},
	volume       = 9,
	pages        = {211--226},
	doi          = {10.1023/A:1008974515925}
}

@article{endsley2017here,
  title={From here to autonomy: lessons learned from human--automation research},
  author={Endsley, Mica R},
  journal={Human factors},
  volume={59},
  number={1},
  pages={5--27},
  year={2017},
  publisher={Sage Publications Sage CA: Los Angeles, CA}, 
  doi = {10.1177/0018720816681350}
}

@article{heyns2018,
  title={Volitional trust, autonomy satisfaction, and engagement at work},
  author={Heyns, Marita and Rothmann, Sebastiaan},
  journal={Psychological reports},
  volume={121},
  number={1},
  pages={112--134},
  year={2018},
  publisher={Sage Publications Sage CA: Los Angeles, CA}, 
    doi = {10.1177/0033294117718555}
}

@article{vandemeulebroucke2018use,
	title        = {The use of care robots in aged care: A systematic review of argument-based ethics literature},
	author       = {Vandemeulebroucke, Tijs and De Casterl{\'e}, Bernadette Dierckx and Gastmans, Chris},
	year         = 2018,
	journal      = {Archives of gerontology and geriatrics},
	publisher    = {Elsevier},
	volume       = 74,
	pages        = {15--25},
	doi          = {10.1016/j.archger.2017.08.014}
}

@article{PRATI2021102072,
	title        = {How to include User eXperience in the design of Human-Robot Interaction},
	author       = {Elisa Prati and Margherita Peruzzini and Marcello Pellicciari and Roberto Raffaeli},
	year         = 2021,
	journal      = {Robotics and Computer-Integrated Manufacturing},
	volume       = 68,
	pages        = 102072,
	doi          = {https://doi.org/10.1016/j.rcim.2020.102072},
	issn         = {0736-5845}
}

@inproceedings{lotz2019you,
	title        = {You’re my mate – acceptance factors for human-robot collaboration in industry},
	author       = {Lotz, V. and Himmel, S. and Ziefle, M.},
	year         = 2019,
	booktitle    = {Proceedings of the {International} {Conference} on {Competitive} {Manufacturing}, COMA'19},
	publisher    = {Department of Industrial Engineering},
	address      = {Stellenbosch, South Africa},
	volume       = 30,
	pages        = {405--411},
	url          = {https://hdl.handle.net/10019.1/105429}
}

@article{zhang2023sorry,
	title        = {“Sorry, it was my fault”: Repairing trust in human-robot interactions},
	author       = {Zhang, Xinyi and Lee, Sun Kyong and Kim, Whani and Hahn, Sowon},
	year         = 2023,
	journal      = {International Journal of Human-Computer Studies},
	publisher    = {Elsevier},
	volume       = 175,
	pages        = 103031,
	doi          = {10.1016/j.ijhcs.2023.103031}
}

@article{lemaignan2016learning,
	title        = {Learning by teaching a robot: The case of handwriting},
	author       = {Lemaignan, S{\'e}verin and Jacq, Alexis and Hood, Deanna and Garcia, Fernando and Paiva, Ana and Dillenbourg, Pierre},
	year         = 2016,
	journal      = {IEEE Robotics \& Automation Magazine},
	publisher    = {IEEE},
	volume       = 23,
	number       = 2,
	pages        = {56--66},
	doi          = {10.1109/MRA.2016.2546700}
}

@article{jung2017exploration,
	title        = {An exploration of the benefits of an animallike robot companion with more advanced touch interaction capabilities for dementia care},
	author       = {Jung, Merel M and Van der Leij, Lisa and Kelders, Saskia M},
	year         = 2017,
	journal      = {Frontiers in ICT},
	publisher    = {Frontiers Media SA},
	volume       = 4,
	pages        = 16,
	doi          = {10.3389/fict.2017.00016}
}

@misc{IEEEEAD,
	title        = {Ethically Aligned Design: A Vision for Prioritizing Human Well-being with Autonomous and Intelligent Systems, Version 2},
	author       = {{The IEEE Global Initiative on Ethics of Autonomous and Intelligent Systems}},
	year         = 2017,
	publisher    = {IEEE},
	url          = {https://standards.ieee.org/wp-content/uploads/import/documents/other/ead_v2.pdf}
}

@book{schneewind1998invention,
	title        = {The invention of autonomy: A history of modern moral philosophy},
	author       = {Schneewind, Jerome B},
	year         = 1998,
	publisher    = {Cambridge University Press}, 
    address = {Cambridge, United Kingdom}
}

@article{guyer2003kant,
	title        = {Kant on the theory and practice of autonomy},
	author       = {Guyer, Paul},
	year         = 2003,
	journal      = {Social philosophy and policy},
	publisher    = {Cambridge University Press},
	volume       = 20,
	number       = 2,
	pages        = {70--98},
	doi          = {10.1017/S026505250320203X}
}

@article{ryan2000self,
	title        = {Self-determination theory and the facilitation of intrinsic motivation, social development, and well-being.},
	author       = {Ryan, Richard M and Deci, Edward L},
	year         = 2000,
	journal      = {American psychologist},
	publisher    = {American Psychological Association},
	volume       = 55,
	number       = 1,
	pages        = {68–--78},
	doi          = {10.1037/0003-066X.55.1.68}
}

@incollection{RYAN2019111,
	title        = {Chapter Four - Brick by Brick: The Origins, Development, and Future of Self-Determination Theory},
	author       = {Richard M. Ryan and Edward L. Deci},
	year         = 2019,
	publisher    = {Elsevier},
	booktitle    = {Advances in Motivation Science},
	volume       = 6,
    address      = {Amsterdam},
	pages        = {111--156},
	doi          = {10.1016/bs.adms.2019.01.001},
	editor       = {Andrew J. Elliot}, 
}

@article{soma2022strengthening,
	title        = {Strengthening human autonomy. In the era of autonomous technology},
	author       = {Soma, Rebekka and Bratteteig, Tone and Saplacan, Diana and Schimmer, Robyn and Campano, Erik and Verne, Guri B},
	year         = 2022,
	journal      = {Scandinavian Journal of Information Systems},
	volume       = 34,
	number       = 2,
	pages        = 5,
	url          = {https://aisel.aisnet.org/sjis/vol34/iss2/5}
}

@article{janssen2024psychological,
	title        = {A psychological need-fulfillment perspective for designing social robots that support well-being},
	author       = {Janssen, Suzanne and Schadenberg, Bob R},
	year         = 2024,
	journal      = {International Journal of Social Robotics},
	publisher    = {Springer},
	volume       = 16,
	number       = 5,
	pages        = {857--878},
	doi          = {10.1007/s12369-024-01102-8}
}

@article{nikolova2024,
	title        = {Robots, meaning, and self-determination},
	author       = {Milena Nikolova and Femke Cnossen and Boris Nikolaev},
	year         = 2024,
	journal      = {Research Policy},
	volume       = 53,
	number       = 5,
	pages        = 104987,
	doi          = {10.1016/j.respol.2024.104987}
}

@article{formosa2021robot,
	title        = {Robot autonomy vs. human autonomy: social robots, artificial intelligence (AI), and the nature of autonomy},
	author       = {Formosa, Paul},
	year         = 2021,
	journal      = {Minds and Machines},
	publisher    = {Springer},
	volume       = 31,
	number       = 4,
	pages        = {595--616},
	doi          = {10.1007/s11023-021-09579-2}
}

@inproceedings{bennett2023,
	title        = {How Does HCI Understand Human Agency and Autonomy?},
	author       = {Bennett, Dan and Metatla, Oussama and Roudaut, Anne and Mekler, Elisa D.},
	year         = 2023,
	booktitle    = {Proceedings of the 2023 {{CHI Conference}} on {{Human Factors}} in {{Computing Systems}}},
	publisher    = {Association for Computing Machinery},
	address      = {New York, USA},
	series       = {{{CHI}} '23},
	pages        = {1--18},
	doi          = {https://doi.org/10.1145/3544548.3580651}
}

@article{peters2018designing,
	title        = {Designing for motivation, engagement and wellbeing in digital experience},
	author       = {Peters, Dorian and Calvo, Rafael A and Ryan, Richard M},
	year         = 2018,
	journal      = {Frontiers in Psychology},
	publisher    = {Frontiers Media SA},
	volume       = 9,
	pages        = 797,
	doi          = {https://doi.org/10.3389/fpsyg.2018.00797}
}

@article{kopp_2024,
	title        = {Facets of Trust and Distrust in Collaborative Robots at the Workplace: Towards a Multidimensional and Relational Conceptualisation},
	author       = {Kopp, Tobias},
	year         = 2024,
	journal      = {International Journal of Social Robotics},
	volume       = 16,
	pages        = {1445--1462},
	doi          = {10.1007/s12369-023-01082-1}
}

@article{simoes2022designing,
	title        = {Designing human-robot collaboration (HRC) workspaces in industrial settings: A systematic literature review},
	author       = {Sim{\~o}es, Ana Correia and Pinto, Ana and Santos, Joana and Pinheiro, Sofia and Romero, David},
	year         = 2022,
	journal      = {Journal of Manufacturing Systems},
	publisher    = {Elsevier},
	volume       = 62,
	pages        = {28--43},
	doi          = {10.1016/j.jmsy.2021.11.007}
}

@inproceedings{donnermann2021,
	title        = {Towards adaptive robotic tutors in universities: A field study},
	author       = {Donnermann, Melissa and Schaper, Philipp and Lugrin, Birgit},
	year         = 2021,
	booktitle    = {International Conference on Persuasive Technology},
	pages        = {33--46},
	doi          = {10.1007/978-3-030-79460-6_3},
    address = {Heidelberg, Germany},
	publisher = {Springer}
}

@article{coghlan2021,
	title        = {Dignity, autonomy, and style of company: dimensions older adults consider for robot companions},
	author       = {Coghlan, Simon and Waycott, Jenny and Lazar, Amanda and Barbosa Neves, Barbara},
	year         = 2021,
	journal      = {Proceedings of the ACM on Human-Computer Interaction},
	publisher    = {Association for Computing Machinery},
	volume       = 5,
	number       = {CSCW1},
	pages        = {1--25},
	doi          = {10.1145/3449178}
}

@inproceedings{ciardo2018,
	title        = {Reduced sense of agency in human-robot interaction},
	author       = {Ciardo, Francesca and De Tommaso, Davide and Beyer, Frederike and Wykowska, Agnieszka},
	year         = 2018,
	booktitle    = {Social Robotics: 10th International Conference, ICSR 2018},
    address = {Qingdao, China},
	pages        = {441--450},
	doi          = {10.1007/978-3-030-05204-1_43},
	publisher = {Springer}
}

@article{barlas2019,
	title        = {When robots tell you what to do: Sense of agency in human-and robot-guided actions},
	author       = {Barlas, Zeynep},
	year         = 2019,
	journal      = {Consciousness and cognition},
	publisher    = {Elsevier},
	volume       = 75,
	pages        = 102819,
	doi          = {10.1016/j.concog.2019.102819}
}

@article{henkemans2017,
	title        = {Design and evaluation of a personal robot playing a self-management education game with children with diabetes type 1},
	author       = {Henkemans, Olivier A Blanson and Bierman, Bert PB and Janssen, Joris and Looije, Rosemarijn and Neerincx, Mark A and van Dooren, Marierose MM and de Vries, Jitske LE and van der Burg, Gert Jan and Huisman, Sasja D},
	year         = 2017,
	journal      = {International Journal of Human-Computer Studies},
	publisher    = {Elsevier},
	volume       = 106,
	pages        = {63--76},
	doi          = {10.1016/j.ijhcs.2017.06.001}
}

@article{huete2011,
	title        = {Personal autonomy rehabilitation in home environments by a portable assistive robot},
	author       = {Huete, Alberto Jard{\'o}n and Victores, Juan G and Martinez, Santiago and Gim{\'e}nez, Antonio and Balaguer, Carlos},
	year         = 2011,
	journal      = {IEEE Transactions on Systems, Man, and Cybernetics, Part C (Applications and Reviews)},
	publisher    = {IEEE},
	volume       = 42,
	number       = 4,
	pages        = {561--570},
	doi          = {10.1109/TSMCC.2011.2159201}
}

@inproceedings{kaur2023,
	title        = {Studying worker perceptions on safety, autonomy, and job security in human-robot collaboration},
	author       = {Kaur, Gurpreet and Banerjee, Sean and Banerjee, Natasha Kholgade},
	year         = 2023,
	booktitle    = {9th International Conference on Automation, Robotics and Applications (ICARA)},
	pages        = {187--191},
	doi          = {10.1109/ICARA56516.2023.10125842},
	publisher = {IEEE}, 
    address = {Abu Dhabi, United Arab Emirates}
}

@article{khalighinejad2016,
	title        = {Social transmission of experience of agency: An experimental study},
	author       = {Khalighinejad, Nima and Bahrami, Bahador and Caspar, Emilie A and Haggard, Patrick},
	year         = 2016,
	journal      = {Frontiers in psychology},
	publisher    = {Frontiers Media SA},
	volume       = 7,
	pages        = 1315,
	doi          = {10.3389/fpsyg.2016.01315}
}

@article{koh2023,
	title        = {The ethics of pet robots in dementia care settings: Care professionals’ and organisational leaders’ ethical intuitions},
	author       = {Koh, Wei Qi and Vandemeulebroucke, Tijs and Gastmans, Chris and Miranda, Rose and Van den Block, Lieve},
	year         = 2023,
	journal      = {Frontiers in psychiatry},
	publisher    = {Frontiers Media SA},
	volume       = 14,
	pages        = 1052889,
	doi          = {10.3389/fpsyt.2023.1052889}
}

@article{lee2023,
	title        = {The role of intrinsic motivations on customers’ service robot use experience},
	author       = {Lee, Seonjeong},
	year         = 2023,
	journal      = {Journal of Quality Assurance in Hospitality \& Tourism},
	publisher    = {Taylor \& Francis},
    volume       = 26,
	number       = 6,
	pages        = {1194-1217},
	doi          = {10.1080/1528008X.2023.2289381}
}

@article{lombardi2023,
	title        = {The impact of facial expression and communicative gaze of a humanoid robot on individual Sense of Agency},
	author       = {Lombardi, Maria and Roselli, Cecilia and Kompatsiari, Kyveli and Rospo, Federico and Natale, Lorenzo and Wykowska, Agnieszka},
	year         = 2023,
	journal      = {Scientific Reports},
	volume       = 13,
	number       = 1,
	pages        = 10113,
	doi          = {10.1038/s41598-023-36864-0}
}

@article{lu2023,
	title        = {Designing social robot for adults using self-determination theory and AI technologies},
	author       = {Lu, Yu and Chen, Chen and Chen, Penghe and Yu, Shengquan},
	year         = 2023,
	journal      = {IEEE Transactions on Learning Technologies},
	publisher    = {IEEE},
	volume       = 16,
	number       = 2,
	pages        = {206--218},
	doi          = {10.1109/TLT.2023.3250465}
}

@article{moradbakhti2023,
	title        = {(Counter-) stereotypical gendering of robots in care: impact on needs satisfaction and gender role concepts in men and women users},
	author       = {Moradbakhti, Laura and Mara, Martina and Castellano, Ginevra and Winkle, Katie},
	year         = 2023,
	journal      = {International Journal of Social Robotics},
	publisher    = {Springer},
	volume       = 15,
	number       = 11,
	pages        = {1769--1790},
	doi          = {doi.org/10.1007/s12369-023-01033-w}
}

@article{ohshima2023,
	title        = {Pocketable-Bones: Self-Augment Mobile Robot Mediating our Sociality},
	author       = {Ohshima, Naoki and Iwasaki, Katsuya and Mayumi, Ryosuke and Hasegawa, Komei and Okada, Michio},
	year         = 2023,
	journal      = {Journal of Robotics and Mechatronics},
	publisher    = {Fuji Technology Press Ltd.},
	volume       = 35,
	number       = 3,
	pages        = {723--733},
	doi          = {10.20965/jrm.2023.p0723}
}

@inproceedings{roselli2019,
	title        = {Robots improve judgments on self-generated actions: an Intentional Binding Study},
	author       = {Roselli, Cecilia and Ciardo, Francesca and Wykowska, Agnieszka},
	year         = 2019,
	booktitle    = {Social Robotics: 11th International Conference, ICSR 2019},
	pages        = {88--97},
	doi          = {10.1007/978-3-030-35888-4_9},
	publisher = {Springer}, 
    address = {Madrid, Spain}
}

@article{sahai2023,
	title        = {Modulations of one’s sense of agency during human--machine interactions: a behavioural study using a full humanoid robot},
	author       = {Saha{\"\i}, A{\"\i}sha and Caspar, Emilie and De Beir, Albert and Grynszpan, Ouriel and Pacherie, Elisabeth and Berberian, Bruno},
	year         = 2023,
	journal      = {Quarterly journal of experimental psychology},
	volume       = 76,
	number       = 3,
	pages        = {606--620},
	doi          = {10.1177/17470218221095841}
}

@article{serholt2022,
	title        = {Comparing a robot tutee to a human tutee in a learning-by-teaching scenario with children},
	author       = {Serholt, Sofia and Ekstr{\"o}m, Sara and K{\"u}ster, Dennis and Ljungblad, Sara and Pareto, Lena},
	year         = 2022,
	journal      = {Frontiers in Robotics and AI},
	publisher    = {Frontiers Media SA},
	volume       = 9,
	pages        = 836462,
	doi          = {10.3389/frobt.2022.836462}
}

@inproceedings{sinai2022,
	title        = {Perceptions of social robots as motivating learning companions for online learning},
	author       = {Sinai, Dafna and Rosenberg-Kima, Rinat B},
	year         = 2022,
	booktitle    = {2022 17th ACM/IEEE International Conference on Human-Robot Interaction (HRI)},
	pages        = {1045--1048},
	doi          = {10.1109/HRI53351.2022.9889592},
	publisher = {IEEE}, 
    address = {Sapporo, Japan}
}

@inproceedings{swift2016,
	title        = {Comparing backstories of a socially assistive robot exercise buddy for adolescent youth},
	author       = {Swift-Spong, Katelyn and Wen, Cheng K Fred and Spruijt-Metz, Donna and Matari{\'c}, Maja J},
	year         = 2016,
	booktitle    = {2016 25th IEEE international symposium on robot and human interactive communication (RO-MAN)},
	pages        = {1013--1018},
	doi          = {10.1109/ROMAN.2016.7745233},
	publisher = {IEEE}, 
    address = {New York, USA}
}

@article{yang2023,
	title        = {Co-creation with service robots and employee wellbeing: a self-determination perspective},
	author       = {Yang, Xue and Gao, Youjiang},
	year         = 2023,
	journal      = {Behaviour \& Information Technology},
	publisher    = {Taylor \& Francis},
    volume = {44},
    number = {10},
	pages        = {2257-2268},
	doi          = {10.1080/0144929X.2023.2295032}
}

@inproceedings{yu2023,
	title        = {Your way or my way: Improving human-robot co-navigation through robot intent and pedestrian prediction visualisations},
	author       = {Yu, Xinyan and Hoggenm{\"u}ller, Marius and Tomitsch, Martin},
	year         = 2023,
	booktitle    = {Proceedings of the 2023 ACM/IEEE International Conference on Human-Robot Interaction},
	pages        = {211--221},
	doi          = {10.1145/3568162.3576992}, 
    publisher = {Association for Computing Machinery}, 
    address = {Stockholm, Sweden}
}

@article{sheldon2008manipulating,
	title        = {Manipulating autonomy, competence, and relatedness support in a game-learning context: New evidence that all three needs matter},
	author       = {Sheldon, Kennon M and Filak, Vincent},
	year         = 2008,
	journal      = {British Journal of Social Psychology},
	publisher    = {Wiley Online Library},
	volume       = 47,
	number       = 2,
	pages        = {267--283},
	doi          = {10.1348/014466607X238797}
}

@article{la2000within,
	title        = {Within-person variation in security of attachment: a self-determination theory perspective on attachment, need fulfillment, and well-being.},
	author       = {La Guardia, Jennifer G and Ryan, Richard M and Couchman, Charles E and Deci, Edward L},
	year         = 2000,
	journal      = {Journal of personality and social psychology},
	publisher    = {American Psychological Association},
	volume       = 79,
	number       = 3,
	pages        = 367,
	doi          = {10.1037/0022-3514.79.3.367}
}

@article{nikou2017mobile,
	title        = {Mobile-Based Assessment: Integrating acceptance and motivational factors into a combined model of Self-Determination Theory and Technology Acceptance},
	author       = {Nikou, Stavros A and Economides, Anastasios A},
	year         = 2017,
	journal      = {Computers in Human Behavior},
	publisher    = {Elsevier},
	volume       = 68,
	pages        = {83--95},
	doi          = {10.1016/j.chb.2016.11.020}
}

@article{lee2018effect,
	title        = {The effect of social networking sites’ activities on customers’ well-being},
	author       = {Lee, Seonjeong},
	year         = 2018,
	journal      = {Journal of Hospitality \& Tourism Research},
	publisher    = {SAGE Publications Sage CA: Los Angeles, CA},
	volume       = 42,
	number       = 7,
	pages        = {1086--1105},
	doi          = {10.1177/1096348016675926}
}

@article{moradbakhti2024development,
	title        = {Development and validation of a basic psychological needs scale for technology use},
	author       = {Moradbakhti, Laura and Leichtmann, Benedikt and Mara, Martina},
	year         = 2024,
	journal      = {Psychological Test Adaptation and Development},
	publisher    = {Hogrefe Publishing},
    volume  = 5,
    pages = {26-45},
	doi          = {10.1027/2698-1866/a000062}
}

@article{nikolova2021perceived,
	title        = {The perceived well-being and health costs of exiting self-employment},
	author       = {Nikolova, Milena and Nikolaev, Boris and Popova, Olga},
	year         = 2021,
	journal      = {Small Business Economics},
	publisher    = {Springer},
	volume       = 57,
	number       = 4,
	pages        = {1819--1836},
	doi          = {10.1007/s11187-020-00374-4}
}

@article{mcauley1989psychometric,
	title        = {Psychometric properties of the Intrinsic Motivation Inventory in a competitive sport setting: A confirmatory factor analysis},
	author       = {McAuley, Edward and Duncan, Terry and Tammen, Vance V},
	year         = 1989,
	journal      = {Research quarterly for exercise and sport},
	publisher    = {Taylor \& Francis},
	volume       = 60,
	number       = 1,
	pages        = {48--58},
	doi          = {10.1080/02701367.1989.10607413}
}

@article{grolnick1989parent,
	title        = {Parent styles associated with children's self-regulation and competence in school.},
	author       = {Grolnick, Wendy S and Ryan, Richard M},
	year         = 1989,
	journal      = {Journal of educational psychology},
	publisher    = {American Psychological Association},
	volume       = 81,
	number       = 2,
	pages        = 143,
	doi          = {10.1037/0022-0663.81.2.143}
}

@article{van2010capturing,
	title        = {Capturing autonomy, competence, and relatedness at work: Construction and initial validation of the Work-related Basic Need Satisfaction scale},
	author       = {Van den Broeck, Anja and Vansteenkiste, Maarten and De Witte, Hans and Soenens, Bart and Lens, Willy},
	year         = 2010,
	journal      = {Journal of occupational and organizational psychology},
	publisher    = {Wiley Online Library},
	volume       = 83,
	number       = 4,
	pages        = {981--1002},
	doi          = {10.1348/096317909X481382}
}

@article{tapal2017sense,
	title        = {The sense of agency scale: A measure of consciously perceived control over one's mind, body, and the immediate environment},
	author       = {Tapal, Adam and Oren, Ela and Dar, Reuven and Eitam, Baruch},
	year         = 2017,
	journal      = {Frontiers in psychology},
	publisher    = {Frontiers Media SA},
	volume       = 8,
	pages        = 1552,
	doi          = {10.3389/fpsyg.2017.01552}
}

@article{MOORE2012546,
	title        = {Intentional binding and the sense of agency: A review},
	author       = {James W. Moore and Sukhvinder S. Obhi},
	year         = 2012,
	journal      = {Consciousness and Cognition},
	volume       = 21,
	number       = 1,
	pages        = {546--561},
	doi          = {10.1016/j.concog.2011.12.002},
}

@book{beauchamp1994principles,
	title        = {Principles of biomedical ethics},
	author       = {Beauchamp, Tom L and Childress, James F},
	year         = 1994,
	publisher    = {Edicoes Loyola}, 
    address = {Oxford, United Kingdom}
    
}

@article{haggard2017sense,
	title        = {Sense of agency in the human brain},
	author       = {Haggard, Patrick},
	year         = 2017,
	journal      = {Nature Reviews Neuroscience},
	publisher    = {Nature Publishing Group UK London},
	volume       = 18,
	number       = 4,
	pages        = {196--207},
	doi          = {10.1038/nrn.2017.14}
}

@article{barlas2018action,
	title        = {Action choice and outcome congruency independently affect intentional binding and feeling of control judgments},
	author       = {Barlas, Zeynep and Kopp, Stefan},
	year         = 2018,
	journal      = {Frontiers in Human Neuroscience},
	publisher    = {Frontiers Media SA},
	volume       = 12,
	pages        = 137,
	doi          = {10.3389/fnhum.2018.00137}
}

@article{perkins2012relational,
	title        = {Relational autonomy in assisted living: A focus on diverse care settings for older adults},
	author       = {Perkins, Molly M and Ball, Mary M and Whittington, Frank J and Hollingsworth, Carole},
	year         = 2012,
	journal      = {Journal of aging studies},
	publisher    = {Elsevier},
	volume       = 26,
	number       = 2,
	pages        = {214--225},
	doi          = {10.1016/j.jaging.2012.01.001}
}

@article{frith2014action,
	title        = {Action, agency and responsibility},
	author       = {Frith, Chris D},
	year         = 2014,
	journal      = {Neuropsychologia},
	publisher    = {Elsevier},
	volume       = 55,
	pages        = {137--142},
	doi          = {10.1016/j.neuropsychologia.2013.09.007}
}

@article{gallagher2000philosophical,
	title        = {Philosophical conceptions of the self: implications for cognitive science},
	author       = {Gallagher, Shaun},
	year         = 2000,
	journal      = {Trends in cognitive sciences},
	publisher    = {Elsevier},
	volume       = 4,
	number       = 1,
	pages        = {14--21},
	doi          = {10.1016/S1364-6613(99)01417-5}
}

@article{moretto2011experience,
	title        = {Experience of agency and sense of responsibility},
	author       = {Moretto, Giovanna and Walsh, Eamonn and Haggard, Patrick},
	year         = 2011,
	journal      = {Consciousness and cognition},
	publisher    = {Elsevier},
	volume       = 20,
	number       = 4,
	pages        = {1847--1854},
	doi          = {10.1016/j.concog.2011.08.014}
}

@article{hackman1976motivation,
	title        = {Motivation through the design of work: Test of a theory},
	author       = {Hackman, J Richard and Oldham, Greg R},
	year         = 1976,
	journal      = {Organizational behavior and human performance},
	publisher    = {Elsevier},
	volume       = 16,
	number       = 2,
	pages        = {250--279},
	doi          = {10.1016/0030-5073(76)90016-7}
}

@article{naneva2020systematic,
	title        = {A systematic review of attitudes, anxiety, acceptance, and trust towards social robots},
	author       = {Naneva, Stanislava and Sarda Gou, Marina and Webb, Thomas L and Prescott, Tony J},
	year         = 2020,
	journal      = {International Journal of Social Robotics},
	publisher    = {Springer},
	volume       = 12,
	number       = 6,
	pages        = {1179--1201},
	doi          = {10.1007/s12369-020-00659-4}
}

@book{EU2024,
	title        = {{ERA} industrial technologies roadmap on human-centric research and innovation for the manufacturing sector},
	author       = {{European Commission and Directorate-General for Research and Innovation}},
	year         = 2024,
	publisher    = {Publications Office of the European Union},
	doi          = {doi/10.2777/0266}, 
    address = {Brussels, Belgium}
}

@misc{europeancommission2025,
	title        = {Horizon {Europe} - Work Programme 2025 Culture, Creativity and Inclusive Society},
	author       = {{European Commission}},
	year         = 2025,
	url          = {https://ec.europa.eu/info/funding-tenders/opportunities/docs/2021-2027/horizon/wp-call/2025/wp-5-culture-creativity-and-inclusive-society_horizon-2025_en.pdf},
	urldate      = {2025-05-20}
}

@misc{europeancommission2025a,
	title        = {Horizon {Europe} - Work Programme 2025 Digital, Industry and Space},
	author       = {{European Commission}},
	year         = 2025,
	url          = {https://ec.europa.eu/info/funding-tenders/opportunities/docs/2021-2027/horizon/wp-call/2025/wp-7-digital-industry-and-space_horizon-2025_en.pdf},
	urldate      = {2025-05-20}
}

@article{turja2022basic,
	title        = {Basic human needs and robotization: How to make deployment of robots worthwhile for everyone?},
	author       = {Turja, Tuuli and S{\"a}rkikoski, Tuomo and Koistinen, Pertti and Melin, Harri},
	year         = 2022,
	journal      = {Technology in Society},
	volume       = 68,
	pages        = 101917,
	doi          = {10.1016/j.techsoc.2022.101917}
}

@inproceedings{dammers2022usability,
	title        = {Usability of human-robot interaction within textile production: Insights into the acceptance of different collaboration types},
	shorttitle   = {Usability of human-robot interaction within textile production},
	author       = {Dammers, Hannah and Vervier, Luisa Sophie and Mittelviefhaus, Lukas and Brauner, Philipp Michael and Ziefle, Martina Cornelia and Gries, Thomas},
	year         = 2022,
	booktitle    = {Usability and User Experience, Vol. 39},
	publisher    = {AHFE Open Access},
	address      = {New York, USA},
	pages        = {213--223},
	doi          = {https://doi.org/10.54941/ahfe1001710}
}

@article{Mori201298,
	title        = {The uncanny valley},
	author       = {Mori, Masahiro and MacDorman, Karl F. and Kageki, Norri},
	year         = 2012,
	journal      = {IEEE Robotics and Automation Magazine},
	volume       = 19,
	number       = 2,
	pages        = {98 – 100},
	doi          = {10.1109/MRA.2012.2192811}
}

@inproceedings{Hegel2008574,
	title        = {Understanding social robots: A user study on anthropomorphism},
	author       = {Hegel, Frank and Krach, Sören and Kircher, Tilo and Wrede, Britta and Sagerer, Gerhard},
	year         = 2008,
	booktitle      = {Proceedings of the 17th IEEE International Symposium on Robot and Human Interactive Communication, RO-MAN},
	pages        = {574 – 579},
	doi          = {10.1109/ROMAN.2008.4600728},
    publisher = {IEEE}, 
    address = {Munich, Germany}
}

@article{DANG2024123683,
	title        = {Viewing machines as humans but humans as machines? Social connectedness shapes the robot anthropomorphism-dehumanization link},
	author       = {Jianning Dang and Li Liu},
	year         = 2024,
	journal      = {Technological Forecasting and Social Change},
	volume       = 208,
	pages        = 123683,
	doi          = {10.1016/j.techfore.2024.123683},
}

@article{bao2023mitigating,
	title        = {Mitigating emotional risks in human-social robot interactions through virtual interactive environment indication},
	author       = {Bao, Aorigele and Zeng, Yi and Lu, Enmeng},
	year         = 2023,
	journal      = {Humanities and Social Sciences Communications},
	publisher    = {Palgrave},
	volume       = 10,
	number       = 1,
	pages        = {1--9},
	doi          = {10.1057/s41599-023-02143-6}
}

@inproceedings{siegel2009persuasive,
	title        = {Persuasive robotics: The influence of robot gender on human behavior},
	author       = {Siegel, Mikey and Breazeal, Cynthia and Norton, Michael I},
	year         = 2009,
	booktitle    = {2009 IEEE/RSJ international conference on intelligent robots and systems},
	pages        = {2563--2568},
	doi          = {10.1109/IROS.2009.5354116},
	publisher = {IEEE}, 
    address = {St. Louis, USA}
}

@article{Roesler2021,
	title        = {A meta-analysis on the effectiveness of anthropomorphism in human-robot interaction},
	author       = {E. Roesler  and D. Manzey  and L. Onnasch},
	year         = 2021,
	journal      = {Science Robotics},
	volume       = 6,
	number       = 58,
	pages        = {eabj5425},
	doi          = {10.1126/scirobotics.abj5425}
}

@misc{europeanparliament2024,
	title        = {Regulation (EU) 2024/1689 of the European Parliament and of the Council},
	author       = {{European Parliament and the Council of the European Union}},
	year         = 2024,
	url          = {https://eur-lex.europa.eu/legal-content/EN/TXT/PDF/?uri=OJ:L_202401689}
}

@inproceedings{yoshikawa2006responsive,
	title        = {Responsive robot gaze to interaction partner.},
	author       = {Yoshikawa, Yuichiro and Shinozawa, Kazuhiko and Ishiguro, Hiroshi and Hagita, Norihiro and Miyamoto, Takanori},
	year         = 2006,
	booktitle    = {Robotics: Science and systems},
	pages        = {37--43},
	doi          = {10.15607/RSS.2006.II.037},
    publisher = {MIT Press}, 
    address = {Philadelphia, USA}
}

@inproceedings{ccakir2023reviewing,
	title        = {Reviewing the social function of eye gaze in social interaction},
	author       = {{\c{C}}ak{\i}r, Mehtap and Huckauf, Anke},
	year         = 2023,
	booktitle    = {Proceedings of the 2023 Symposium on Eye Tracking Research and Applications},
	pages        = {1--3},
	doi          = {10.1145/3588015.3589513},
    publisher = {Association for Computing Machinery}, 
    address = {Tubingen, Germany}
}

@article{szalma2009individual,
	title        = {Individual differences in human--technology interaction: Incorporating variation in human characteristics into human factors and ergonomics research and design},
	author       = {Szalma, James L},
	year         = 2009,
	journal      = {Theoretical Issues in Ergonomics Science},
	publisher    = {Taylor \& Francis},
	volume       = 10,
	number       = 5,
	pages        = {381--397},
	doi          = {10.1080/14639220902893613}
}

@article{venkatesh2012consumer,
	title        = {Consumer acceptance and use of information technology: extending the unified theory of acceptance and use of technology},
	author       = {Venkatesh, Viswanath and Thong, James YL and Xu, Xin},
	year         = 2012,
	journal      = {MIS quarterly},
	publisher    = {JSTOR},
	pages        = {157--178},
	doi          = {10.2307/41410412}
}

@article{anglim2020predicting,
	title        = {Predicting psychological and subjective well-being from personality: A meta-analysis.},
	author       = {Anglim, Jeromy and Horwood, Sharon and Smillie, Luke D and Marrero, Rosario J and Wood, Joshua K},
	year         = 2020,
	journal      = {Psychological bulletin},
	publisher    = {American Psychological Association},
	volume       = 146,
	number       = 4,
	pages        = 279,
	doi          = {10.1037/bul0000226}
}

@article{koestner2023causality,
	title        = {Causality orientations theory: SDT’s forgotten mini-theory},
	author       = {Koestner, Richard and Levine, Shelby L},
	year         = 2023,
	journal      = {The Oxford handbook of self-determination theory},
	publisher    = {Oxford University Press},
	pages        = {124--138},
	doi          = {10.1093/oxfordhb/9780197600047.001.0001}
}

@article{cummins2014agency,
	title        = {Agency is distinct from autonomy},
	author       = {Cummins, Fred},
	year         = 2014,
	journal      = {AVANT. Pismo Awangardy Filozoficzno-Naukowej},
	publisher    = {O{\'s}rodek Bada{\'n} Filozoficznych},
	number       = 2,
	pages        = {98--112}
}

@article{moore2016sense,
	title        = {What is the sense of agency and why does it matter?},
	author       = {Moore, James W},
	year         = 2016,
	journal      = {Frontiers in psychology},
	publisher    = {Frontiers Media SA},
	volume       = 7,
	pages        = 1272,
	doi          = {10.3389/fpsyg.2016.01272}
}

@article{antusch2021intentional,
	title        = {Intentional action and limitation of personal autonomy. Do restrictions of action selection decrease the sense of agency?},
	author       = {Antusch, S and Custers, R and Marien, H and Aarts, H},
	year         = 2021,
	journal      = {Consciousness and Cognition},
	publisher    = {Elsevier},
	volume       = 88,
	pages        = 103076,
	doi          = {10.1016/j.concog.2021.103076}
}

@article{wegner1999apparent,
	title        = {Apparent mental causation: Sources of the experience of will.},
	author       = {Wegner, Daniel M and Wheatley, Thalia},
	year         = 1999,
	journal      = {American psychologist},
	publisher    = {American Psychological Association},
	volume       = 54,
	number       = 7,
	pages        = 480,
	doi          = {10.1037/0003-066X.54.7.480}
}

@misc{elsevier2025,
	title        = {ScienceDirect: The premier platform for scientific, health and technical literature},
	author       = {Elsevier},
	year         = {2025},
	publisher    = {Elsevier},
	url          = {https://www.elsevier.com/products/sciencedirect}
}

@misc{elsevier2025a,
	title        = {Scopus: A comprehensive abstract and citation database for impact makers},
	author       = {Elsevier},
	year         = {2025},
	publisher    = {Elsevier},
	url          = {https://www.elsevier.com/products/scopus}
}

@misc{clarivate2025,
	title        = {Web of Science platform},
	author       = {Clarivate},
	year         = {2025},
	publisher    = {Clarivate},
	url          = {https://clarivate.com/academia-government/scientific-and-academic-research/research-discovery-and-referencing/web-of-science/}
}

@misc{IEEE2025,
	title        = {About IEEE Xplore},
	author       = {IEEE},
	year         = {2025},
	publisher    = {IEEE},
	url          = {https://ieeexplore.ieee.org/Xplorehelp/overview-of-ieee-xplore/about-ieee-xplore}
}

@misc{ACM2025,
	title        = {About the ACM Digital Library},
	author       = {ACM},
	year         = {2025},
	publisher    = {ACM},
	url          = {https://dl.acm.org/about}
}

@article{chen2015basic,
  title={Basic psychological need satisfaction, need frustration, and need strength across four cultures},
  author={Chen, Beiwen and Vansteenkiste, Maarten and Beyers, Wim and Boone, Liesbet and Deci, Edward L and Van der Kaap-Deeder, Jolene and Duriez, Bart and Lens, Willy and Matos, Lennia and Mouratidis, Athanasios and others},
  journal={Motivation and emotion},
  volume={39},
  number={2},
  pages={216--236},
  year={2015},
  publisher={Springer}, 
doi = {10.1007/S11031-014-9450-1}
}

@article{sheldon2009psychological,
  title={Psychological needs as basic motives, not just experiential requirements},
  author={Sheldon, Kennon M and Gunz, Alexander},
  journal={Journal of personality},
  volume={77},
  number={5},
  pages={1467--1492},
  year={2009},
  publisher={Wiley Online Library}, 
    doi = {10.1111/j.1467-6494.2009.00589.x}
}

@article{BALTAZAR2014120,
title = {Eye contact elicits bodily self-awareness in human adults},
journal = {Cognition},
volume = {133},
number = {1},
pages = {120-127},
year = {2014},
issn = {0010-0277},
doi = {https://doi.org/10.1016/j.cognition.2014.06.009},
author = {Matias Baltazar and Nesrine Hazem and Emma Vilarem and Virginie Beaucousin and Jean-Luc Picq and Laurence Conty},
}

@article{ULLOA2019102794,
title = {The impact of eye contact on the sense of agency},
journal = {Consciousness and Cognition},
volume = {74},
pages = {102794},
year = {2019},
issn = {1053-8100},
doi = {https://doi.org/10.1016/j.concog.2019.102794},
author = {José Luis Ulloa and Roberta Vastano and Nathalie George and Marcel Brass},
}

@article{CONTY2016184,
title = {Watching Eyes effects: When others meet the self},
journal = {Consciousness and Cognition},
volume = {45},
pages = {184-197},
year = {2016},
issn = {1053-8100},
doi = {https://doi.org/10.1016/j.concog.2016.08.016},
author = {Laurence Conty and Nathalie George and Jari K. Hietanen},
}

@article{YOSHIE20132028,
title = {Negative Emotional Outcomes Attenuate Sense of Agency over Voluntary Actions},
journal = {Current Biology},
volume = {23},
number = {20},
pages = {2028-2032},
year = {2013},
issn = {0960-9822},
doi = {https://doi.org/10.1016/j.cub.2013.08.034},
author = {Michiko Yoshie and Patrick Haggard},
}

@article{greenberg1992depression,
  title={Depression, self-focused attention, and the self-serving attributional bias},
  author={Greenberg, Jeff and Pyszczynski, Tom and Burling, John and Tibbs, Karyn},
  journal={Personality and Individual Differences},
  volume={13},
  number={9},
  pages={959--965},
  year={1992},
  publisher={Elsevier}, 
  doi = {10.1016/0191-8869(92)90129-D}
}

@inproceedings{bloch2023comparison,
  title={Comparison of Attitudes Towards Robots of Different Population Samples in Norway},
  author={Bloch, Marten and Fernandes, Alexandra},
  booktitle={Companion of the 2023 ACM/IEEE International Conference on Human-Robot Interaction},
  pages={576--579},
  year={2023}, 
  publisher = {Association for Computing Machinery},
  address = {Stockholm, Sweden},
  doi = {10.1145/3568294.3580151}  
}

@inproceedings{nikolovska2024user,
  title={User perception of robot behavior as a function of previous experience with robots},
  author={Nikolovska, Kristina and Pohl, Jan and Hommel, Bernhard and Kappas, Arvid and Maurelli, Francesco},
  booktitle={2024 16th International Conference on Human System Interaction (HSI)},
  pages={1--7},
  year={2024},
  organization={IEEE}, 
  address = {Paris, France}, 
  doi = {10.1109/HSI61632.2024.10613554}
}

@inproceedings{coyle2012did,
  title={I did that! Measuring users' experience of agency in their own actions},
  author={Coyle, David and Moore, James and Kristensson, Per Ola and Fletcher, Paul and Blackwell, Alan},
  booktitle={Proceedings of the SIGCHI conference on human factors in computing systems},
  pages={2025--2034},
  year={2012}, 
  doi = {10.1145/2207676.2208350}, 
  publisher = {Association for Computing Machinery}, 
  address = {Austin Texas, USA}
}

@article{salthouse2003needs,
  title={What needs to be explained to account for age-related effects on multiple cognitive variables?},
  author={Salthouse, Timothy A and Ferrer-Caja, Emilio},
  journal={Psychology and aging},
  volume={18},
  number={1},
  pages={91},
  year={2003},
  publisher={American Psychological Association}, 
  doi = {10.1037/0882-7974.18.1.91}
}

@article{czaja2006factors,
  title={Factors predicting the use of technology: findings from the Center for Research and Education on Aging and Technology Enhancement {(CREATE)}.},
  author={Czaja, Sara J and Charness, Neil and Fisk, Arthur D and Hertzog, Christopher and Nair, Sankaran N and Rogers, Wendy A and Sharit, Joseph},
  journal={Psychology and aging},
  volume={21},
  number={2},
  pages={333–352},
  year={2006},
  publisher={American Psychological Association}, 
  doi = {10.1037/0882-7974.21.2.333}
}

@article{ueda2021influence,
  title={Influence of levels of automation on the sense of agency during continuous action},
  author={Ueda, Sayako and Nakashima, Ryoichi and Kumada, Takatsune},
  journal={Scientific reports},
  volume={11},
  number={1},
  pages={2436},
  year={2021},
  publisher={Nature Publishing Group UK London}, 
  doi = {10.1038/s41598-021-82036-3}
}

@article{mariano2024aging,
  title={How aging shapes our sense of agency},
  author={Mariano, Marika and Kuster, Nicole and Tartufoli, Matilde and Zapparoli, Laura},
  journal={Psychonomic Bulletin \& Review},
  volume={31},
  number={4},
  pages={1714--1722},
  year={2024},
  publisher={Springer}, 
  doi = {10.3758/s13423-023-02449-1}
}

@article{he2022technology,
  title={Technology acceptance in socially assistive robots: Scoping review of models, measurement, and influencing factors},
  author={He, Ying and He, Qiu and Liu, Qian},
  journal={Journal of Healthcare Engineering},
  volume={2022},
  number={1},
  pages={6334732},
  year={2022},
  publisher={Wiley Online Library}, 
  doi = {10.1155/2022/6334732}
}

@article{concetta2025toward,
  title={Toward acceptance of human-robot collaboration in industrial settings: a bibliometric and systematic literature review},
  author={Concetta Manuela, La Fata and Antonio, Giallanza and Giada, La Scalia and Rosa, Micale and others},
  journal={The International Journal of Advanced Manufacturing Technology},
  pages={2139–2160},
  volumne = {139}, 
  year={2025},
  publisher={Springer}, 
  doi = {10.1007/s00170-025-16039-z}
}

@article{DEGRAAF20131476,
title = {Exploring influencing variables for the acceptance of social robots},
journal = {Robotics and Autonomous Systems},
volume = {61},
number = {12},
pages = {1476-1486},
year = {2013},
issn = {0921-8890},
doi = {https://doi.org/10.1016/j.robot.2013.07.007},
author = {Maartje M.A. {de Graaf} and Somaya {Ben Allouch}},
}

@book{ryan2018,
	address = {New York London},
	edition = {Paperback edition},
	series = {Psychology},
	title = {Self-determination theory: basic psychological needs in motivation, development, and wellness},
	publisher = {The Guilford Press},
	author = {Ryan, Richard M. and Deci, Edward L.},
	year = {2018},
	doi = {10.1521/978.14625/28806}
}

@article{ryan2022we,
  title={We know this much is (meta-analytically) true: A meta-review of meta-analytic findings evaluating self-determination theory.},
  author={Ryan, Richard M and Duineveld, Jasper J and Di Domenico, Stefano I and Ryan, William S and Steward, Ben A and Bradshaw, Emma L},
  journal={Psychological Bulletin},
  volume={148},
  number={11-12},
  pages={813-842},
  year={2022},
  publisher={American Psychological Association}, 
  doi = {10.1037/bul0000385}
}

@article{stanley2021meta,
  title={A meta-analytic investigation of the relationship between basic psychological need satisfaction and affect.},
  author={Stanley, Peter J and Schutte, Nicola S and Phillips, Wendy J},
  journal={Journal of Positive School Psychology},
  volume={5},
  number={1},
  year={2021}, 
  pages={1–16},
  doi = {10.47602/jpsp.v5i1.210}  
}

@article{baard2004intrinsic,
  title={Intrinsic need satisfaction: a motivational basis of performance and weil-being in two work settings},
  author={Baard, Paul P and Deci, Edward L and Ryan, Richard M},
  journal={Journal of applied social psychology},
  volume={34},
  number={10},
  pages={2045--2068},
  year={2004},
  publisher={Wiley Online Library}, 
  doi = {10.1111/j.1559-1816.2004.tb02690.x}
}

@inproceedings{sauppe2015social,
  title={The social impact of a robot co-worker in industrial settings},
  author={Saupp{\'e}, Allison and Mutlu, Bilge},
  booktitle={Proceedings of the 33rd annual ACM conference on human factors in computing systems},
  pages={3613--3622},
  year={2015}, 
    doi = {10.1145/2702123.2702181}, 
  publisher = {Association for Computing Machinery}, 
  address = {Seoul, Republic of Korea}
    
}

\end{document}